\documentstyle[12pt,epsfig]{article}
\voffset     = - 0.10 in
\begin{document}
\thispagestyle{empty}


{\huge\bf
\begin{center}
Study of the Hadron Shower Profiles with the Tile Hadron Calorimeter 
\end{center}
}

\vspace*{\fill}

\begin{center}

{\Large\bf
Y.A.~Kulchitsky, V.S.~Rumyantsev, \\
}
{\it
Institute of Physics, Academy of Sciences, Minsk, Belarus \\
\& JINR, Dubna, Russia
}

{\Large\bf
J.A.~Budagov, N.A.~Russakovich, V.B.~Vinogradov \\
}
{\it
JINR, Dubna, Russia
}

{\Large\bf
M.~Nessi \\
}
{\it
CERN, Geneva, Switzerland
}
\end{center}

\vspace*{\fill}

\begin{abstract}
The lateral and longitudinal profiles of the hadronic showers detected by
iron-scintillator tile hadron calorimeter with
longitudinal tile configuration have been investigated.
The results are based on $100\ GeV$ pion beam data.
Due to the beam scan provided many different beam
impact locations with cells it is
succeeded to obtain detailed picture of transverse shower behavior.
The underlying radial energy densities for four depths and for 
overall calorimeter have  been reconstructed.
The three-dimensional hadronic shower parametrisation
have been suggested.
\end{abstract}

\newpage

{


\section{Introduction}

Hadronic shower is a basic notion of hadron calorimetry.
But despite  that hadronic shower characteristics are studied 
for many years the exhaustive quantitative understanding
of hadronic shower properties is not exist.
The published data are as a rule the energy deposition in calorimeter
cells and therefore are related with specific cell
dimensions and the acceptance
of cells relative to shower axis.
Furthermore,  as to the transverse profiles
they are as usual the energy depositions
as a function of transverse coordinates,  not a radius,  and integrated over
the other coordinate
\cite{womersley88}.
Meanwhile for many purposes of experiments 
a very detailed simulation is not needed and
a three dimensional parameterisation of hadron shower
development is to become very important for fast simulation which
significantly (up to $10^5$ times) to speed a 
detailed GEANT based simulation
\cite{bock81},
\cite{grindhammer90},
\cite{brun91}.


In this paper we report on the results of the experimental study
of hadronic shower profiles
detected by prototype of ATLAS tile hadron calorimeter
\cite{atcol94}.
This calorimeter has innovative concept of longitudinal
segmentation of active and passive layers (see
Fig.~\ref{fig:f1}) and
the measurement of hadron shower profiles therefore a special interest
\cite{lokajicek95-63}.
This investigation was performed on the basis of data from 100 GeV pion
exposure of the prototype calorimeter
at the CERN SPS
at different $z$ impact points
in the range from $- 36$ to $20$ cm
($z$ scan) at incident angle $\Theta = 10^{o}$ which were obtained
in May 1995.

Earlier some results related with lateral shower profiles for this 
calori\-meter  were obtained in
\cite{juste95}.

\begin{figure*}[tbph]
     \begin{center}
        \begin{tabular}{|c|}
        \hline
      \mbox{\epsfig{figure=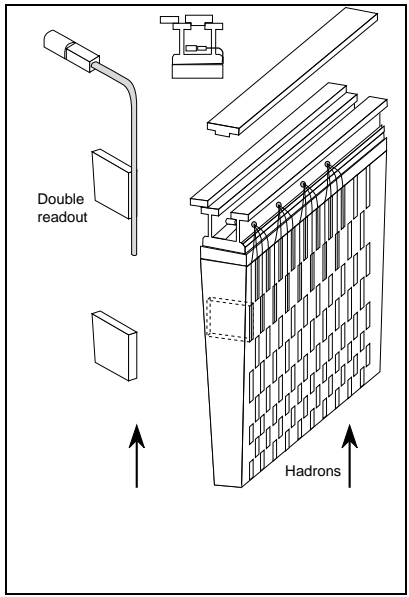,width=0.25\textheight}}\\
        \hline
        \end{tabular}
     \end{center}
       \caption{
       Principle of the tile hadronic calorimeter.
       \label{fig:f1}}
\end{figure*}

\section{The  Calorimeter}

The  prototype the ATLAS hadron tile calorimeter (Fig.\ \ref{fig:f01})
is composed of five  modules.
Each module 
spans $2 \pi / 64$ in azimuthal angle,  100 cm in the $z$ direction,
180 cm in the radial direction
(about 9 interaction lengths $\lambda_I$ at $\eta = 0$ or to 
about 80 effective radiation length $X_o$),
and has  a front face of 100 $\times$ 20 cm$^2$
\cite{berger95}.
The iron structure of each module consists of 57 repeated "periods".
Each period is 18~mm thick and consists of four layers.
The first and third layers are formed by large trapezoidal steel plates
(master plates),  5 mm thick and spanning the full radial dimension of the
module.
In the second and fourth layers,  smaller trapezoidal steel plates
(spacer plates) and scintillator tiles alternate along the radial direction.
These layers consist of 18 different trapezoids of steel or scintillator,
each spanning 100 mm along $x$ depending on their radial position.
The spacer plates and scintillator tiles are 4 mm and 3 mm thick
respectively.
The iron to scintillator ratio is $4.67 : 1$ by volume.
The calorimeter thickness along $x$ direction at incidence angle
$\Theta = 10^{o}$ corresponds to 1.49 m of iron equivalent
\cite{lokajicek95-64}.

\begin{figure*}[tbph]
     \begin{center}
        \begin{tabular}{|c|}
        \hline
      \mbox{\epsfig{figure=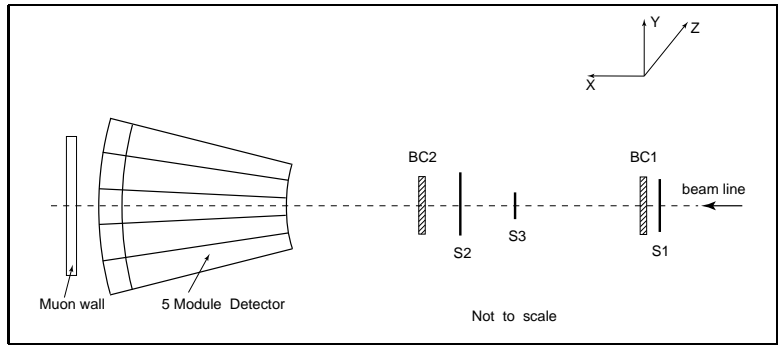,width=0.50\textheight}}\\
        \hline
        \end{tabular}
     \end{center}
       \caption{
       Setup of the tile hadronic calorimeter prototype.
       \label{fig:f01}}
\end{figure*}

Radially oriented WLS fibers collect light from the tiles at both of
their open edges and bring it to photo-multipliers (PMTs) at the periphery
of the calorimeter.
Each PMT views a specific group of tiles,  through the corresponding
bundle of fibers.
The calorimeter is radially segmented into four depth segments by
grouping fibers from different tiles.
As a result of  each module is divided on 
5 (along $z$) $\times$ 4 (along $x$) separate cells.

The readout cells have lateral dimensions of 
200 mm (along $z$) $\times$ ($200 \div 300$) mm 
(along $y$, depending from a depth number) 
and longitudinal dimensions of 300, 400, 500, 600 mm for depths 1-4,
corresponding to 1.5, 2, 2.5 and 3 $\lambda_{I}$ at $\eta=0$.
On the output we have for each event 200 values of energies
$E_{i,j,k,l}$
from $PMT$ properly calibrated
\cite{berger95}
with pedestal subtracted.
Here indexes i, j, k, l mean: $i = 1, \ldots, 5$ is the column of cells number
(tower),
$j = 1, \ldots, 5$ is the row (module) number, 
$k = 1, \ldots, 4$ is the depth number and $l = 1, 2$ is the $PMT$ number.

The calorimeter has been positioned on a scanning table,  able to allow
high precision movements along any direction.
Upstream of the calorimeter,  a trigger counter telescope was installed,
defining a beam spot of 2~cm diameter.
Two delay-line wire chambers,  each with $(z, y)$  readout,
allowed to reconstruct the impact point of beam particles on the
calorimeter face to better than $\pm$~1~mm
\cite{ariztizabal94}.
For the measurements of the hadronic shower longitudinal and lateral leakages
backward (80$\times$80 cm$^2$) and side (40$\times$115 cm$^2$)
``muon walls'' were placed
behind and side the calorimeter modules
\cite{lokajicek95-63}.

Construction and performance of ATLAS
iron-scintillator barrel ha\-d\-ron prototype calorimeter is described
elsewhere
\cite{atcol94},
\cite{berger95},
\cite{budagov96-72},
\cite{tilecal-tdr96}.

\section{The methods for extracting the underlying radial energy density}

The incident particle interacts with the material of calorimeter and
a shower is initiated.
In the following the used coordinate system will be based on the 
hadron shower direction.
The axis of the shower, defined to be a track of a incident particle,
forms the $x$ axis with $x=0$ at the calorimeter front face.

We measure the energy depositions in calorimetry cells.
In $ijk$-cell of the calorimeter with volume $V_{ijk}$ and cell 
center coordinates 
$(x, y, z)$ the energy deposition $E_{ijk}$ is
\begin{equation}
E_{ijk} (x, y, z) =
\int
\int_{V_{ijk}}
\int
f(x,y,z)dxdydz,
\label{e010}
\end{equation}
where 
f(x,y,z) 
is the three-dimensional hadron shower energy density function.

In the following we will consider the various marginal (integrated) densities 
\cite{review96}:
longitudinal density $f_1 (x)$ obtained after integrating $f(x,y,z)$
over $y$ and $z$ coordinates
\begin{equation}
f_1 (x) =
        \int_{y_{min}}^{y_{max}}
        \int_{z_{min}}^{z_{max}}
        f(x,y,z)dydz , 
\label{e08}
\end{equation}
transverse densities $f_{2} (y,z)$ for four
depths and overall calorimeter 
obtained after integrating  $f (x,y,z)$ over the various $x$ ranges
\begin{equation}
f_{2} (y,z) = \int_{x_{1}}^{x_{2}} f(x,y,z) dx ,
\label{e09}
\end{equation}
transverse densities $f  (z)$
\begin{equation}
f  (z) = 
\int_{x_{1}}^{x_{2}} 
\int_{ - \infty}^{\infty}  
f(x,y,z) dx dy .
\label{e09-1}
\end{equation}
Cumulative function is 
\begin{equation}
F (z) = \int_{-\infty}^{z} \int_{ - \infty}^{\infty} f_{2} (z, y) dz dy
\label{e3-01}
\end{equation}
and related with marginal density $f  (z)$ by  
\begin{equation}
f  (z) = \frac{d F(z)}{dz}.
\label{e30-01}
\end{equation}

Due to the azimuthal symmetry densities $f_{2} (y, z)$ are
the function only a radius
$r =  \sqrt(y^2+z^2)$ from the shower axis, i.e.\
\begin{equation}
f_{2} (y,z) = \Phi (r).
\label{e30-02}
\end{equation}
In such case 
\begin{equation}
E_{ijk} = 
\int \int_{V_{ijk}} \Phi (r) r dr d\phi ,
\label{e3-03}
\end{equation}
where $\phi$ is the azimuthal angle.

There are some methods for extracting of radial density $\Phi(r)$ 
from the measured distributions of energy depositions $E_{ijk}$.

One method is the unfolding $\Phi(r)$ from (\ref{e3-03}).
This method was used in the analysis of data from
the lead-scintillating fiber Spaghetti Calorimeter
\cite{acosta92}.
Several analytic forms of $\Phi (r)$  were tried,
but the simplest that describes the 
energy deposition in cells was a combination of an exponential and a Gaussian:
\begin{equation}
\Phi (r) = \frac{a_1}{r} e^{ - \frac{r}{\lambda_1}} +
           \frac{a_2}{r} e^{ - {( \frac{r}{\lambda_{2}} )}^2 }.
\label{e1}
\end{equation}
In order to determine the free parameters $a_i, \lambda_i$ in the expression
(\ref{e1}) a $\chi^2$ minimisation fit have been done.

Another  method is using the marginal density function $f  (z)$
and its connection with radial density $\Phi (r)$
\cite{lednev95}.
\begin{equation}
f  (z) = 2 \int_{|z|}^{\infty} \frac{\Phi(r) r dr}{\sqrt(r^{2}-z^{2})} .
\label{e4}
\end{equation}
This method was used \cite{lednev95} 
for extracting of electron shower transverse profile
on the basis of the data from $GAMS-2000$ electromagnetic calorimeter 
\cite{akopdijanov77}.

Integral equation (\ref{e4}) can be reduced to the Abelian equation
by replacing of variables 
\cite{whitteker27}.
We solved equation (\ref{e4}) (see Appendix 1) and obtained
\begin{equation}
\Phi(r) = - \frac{1}{\pi} \frac{d}{dr^2}
\int_{r^2}^{\infty} \frac{f  (z) d z^2}{\sqrt(z^{2}-r^{2})}.
\label{e5}
\end{equation}

The marginal density $f  (z)$ also may be determined by various approaches.
One approach is the using of cumulative function $f  (z)$
\cite{lednev95}
and differentiation of it according to (\ref{e30-01}).
Another approach is to assume some form of marginal density $f  (z)$, 
to derive a formula for energy deposition in tower and
to determine  parameters of $f  (z)$ by fitting 
\cite{gavrish91}.


If the three exponential distribution for parametrisation of 
$f  (z)$ is used  
\cite{binon83},
\cite{gavrish91}:
\begin{equation}
f  (z)  = 
 \frac{E_o}{2B}  \sum_{i=1}^{3}  a_i 
e^{ - \frac{|z|}{{\lambda}_{i}} }, 
\label{e21}
\end{equation}
then for the energy deposition in a tower $E (z)$,
the cumulative function $F (z)$, for radial density $\Phi (r)$ 
we obtain:
\begin{equation}
E (z) = 
\int_{z - h/2}^{z + h/2}
f  ( z^{\prime} - z ) dz^{\prime},
\label{e22-01}
\end{equation}

\begin{eqnarray}
E ( z)  = 
&  \frac{E_{o}}{B}  \sum_{i=1}^{3}  a_i {\lambda}_i
( 1 - e^{ - \frac{h}{2 {\lambda}_i} } 
ch( \frac{ |z| }{ {\lambda}_i })),
&  for\    |z| \leq \frac{h}{2},
\\
E ( z ) =  
& \frac{E_o}{B} \sum_{i = 1}^{3} a_i {\lambda}_i
sh(\frac{h}{2 \lambda_i }) \cdot 
e^{ - \frac{|z|}{{\lambda}_i}},
&  for\   | z | \geq \frac{h}{2},
\label{e023}
\end{eqnarray}

\begin{eqnarray}
F (z_r) = 
& \frac{E_o}{2B} \sum_{i=1}^{3}  a_i \lambda_i
e^{ \frac{ z_r }{ \lambda_{i} } },  
& for\ z_r \leq 0,
\\
F (z_r) = 
& \frac{E_o}{2} + \frac{E_o}{2B} \sum_{i=1}^{3}  a_i \lambda_i
(1 - e^{- \frac{z_r}{\lambda_{i}}}), 
& for\ z_r \geq 0,
\label{e23-2}
\end{eqnarray}

\begin{equation}
\Phi (r) = \frac{E_o}{2B} \sum_{i=1}^{3} \frac{a_{i}}{\pi \lambda_{i}}
K_{0} (\frac{r}{\lambda_{i}}), 
\label{e23}
\end{equation}
where 
$z$ is the transverse coordinate of the tower center,
$z_r$ is the right edge transverse coordinate of the tower in the center,
$h$ is the front face size of towers along $Z$ axis
($h = 200 \ mm$ in our case),
$B =  \sum_{i=1}^{3} a_i \lambda_i$, 
$E_o,\ a_i,\ \lambda_i$ are free parameters,
$E_o$ is an energy normalisation factor,
$K_{0}$ is modified Bessel function.
Using the condition $\sum_{i=1}^{3} a_i = 1$
we reduced the number of parameters to six.

The radial containment of shower as a function of $r$ is
\begin{equation}
I (r) = 
\int_0^r \int_0^{2 \pi} \Phi (r) r dr d\phi =
\frac{E_o}{B} \sum_{i=1}^{3} a_{i} {\lambda}_{i} 
(1 - \frac{r}{{\lambda}_{i}} K_1(\frac{r}{{\lambda}_{i}})),
\label{e008}
\end{equation}
where $K_{1}$ is modified Bessel function.

\section{Results}

Using the program TILEMON
\cite{efthymiopoulos95} 
30 runs with different impact points of incident particles 
in the range from $- 36$ to $20$ cm
contained 320~K events have been analysed
and the various energy deposition spectra have been obtained.

\subsection{Transverse behaviour of hadron showers}

\subsubsection{Energy deposition in the towers}

\begin{figure*}[tbph]
     \begin{center}
        \begin{tabular}{|c||c|}
        \hline
        \mbox{\epsfig{figure=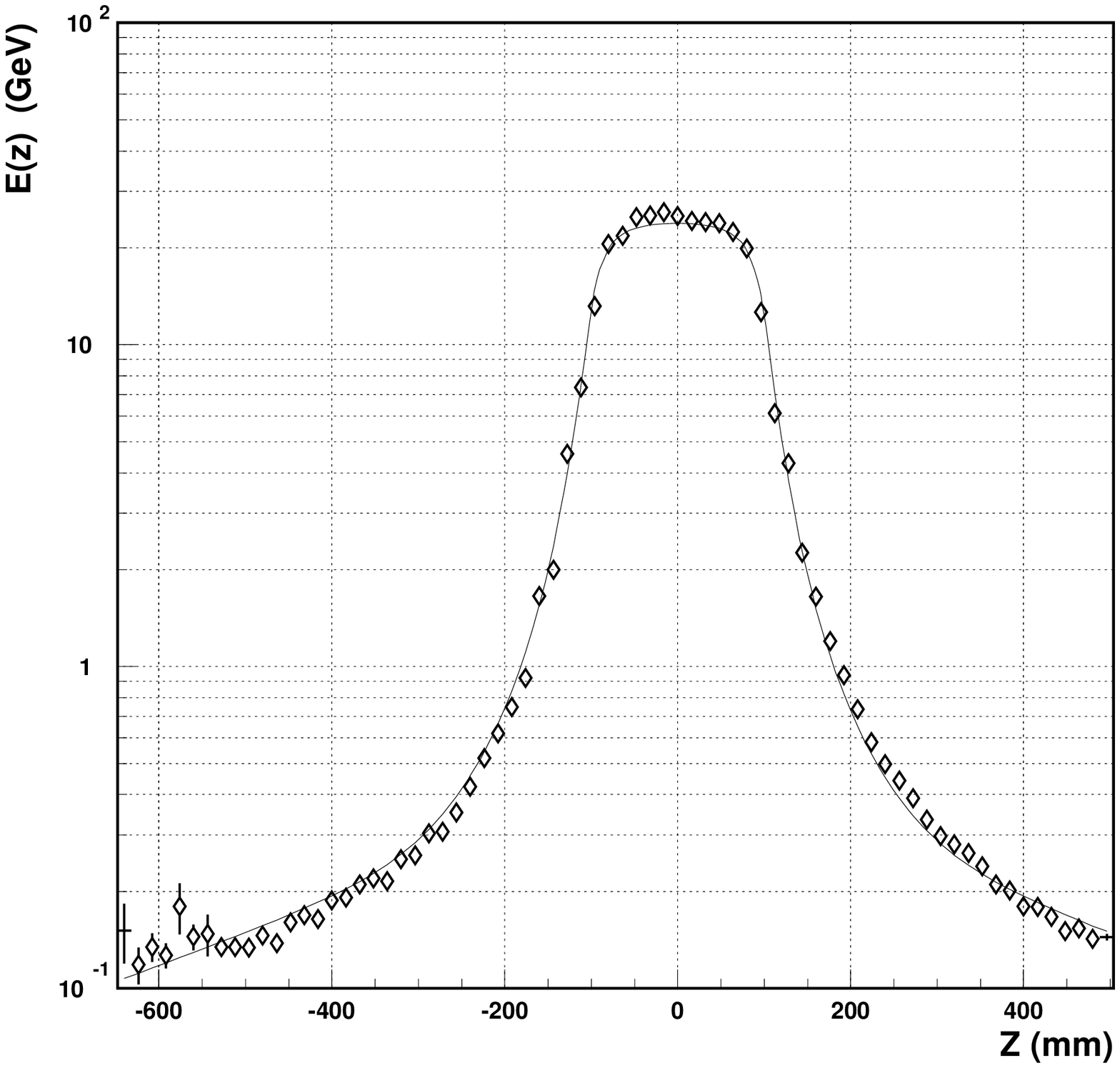,width=0.30\textheight}}
        &
        \mbox{\epsfig{figure=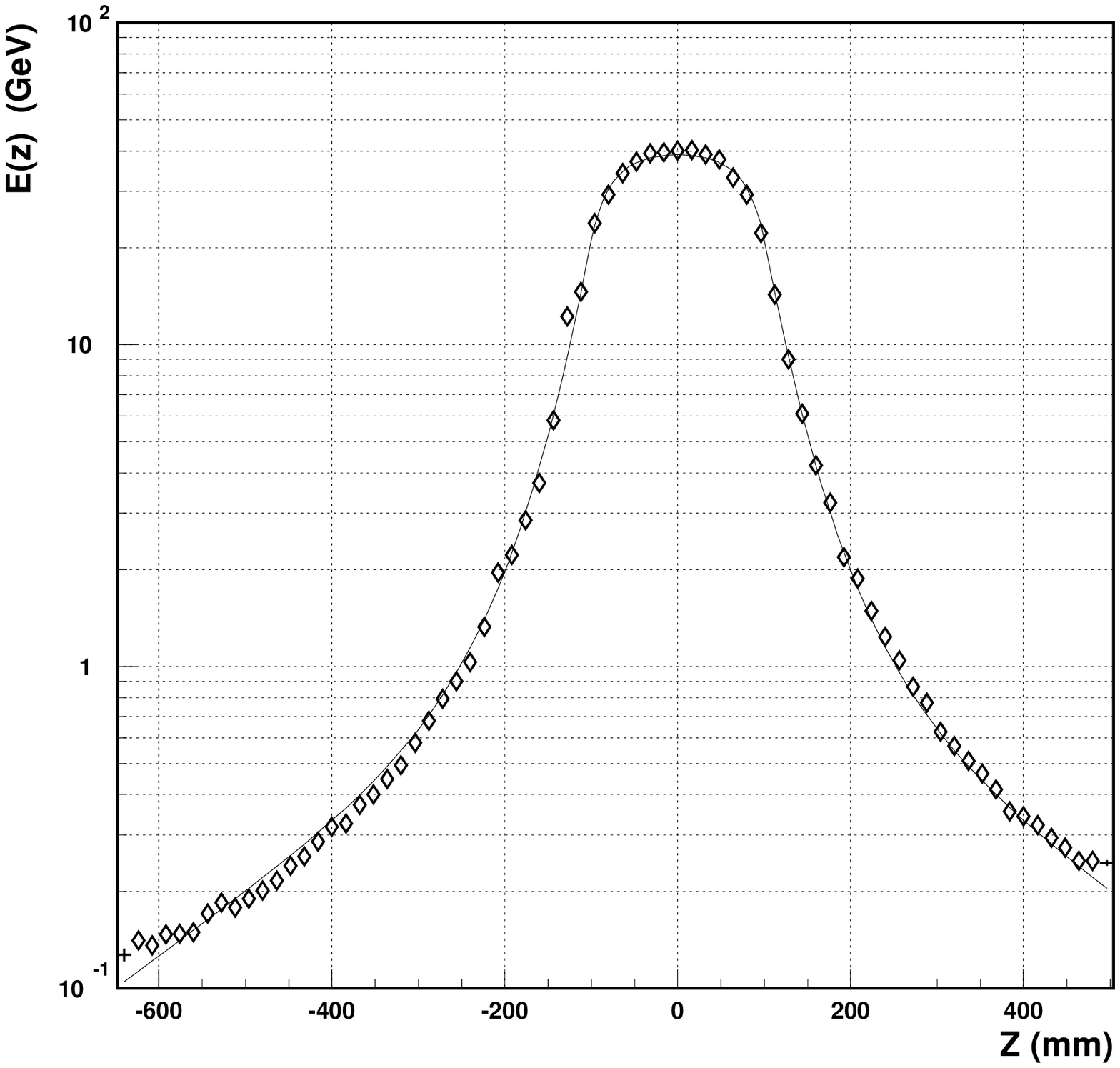,width=0.30\textheight}}
        \\
        \hline
        \hline
        \mbox{\epsfig{figure=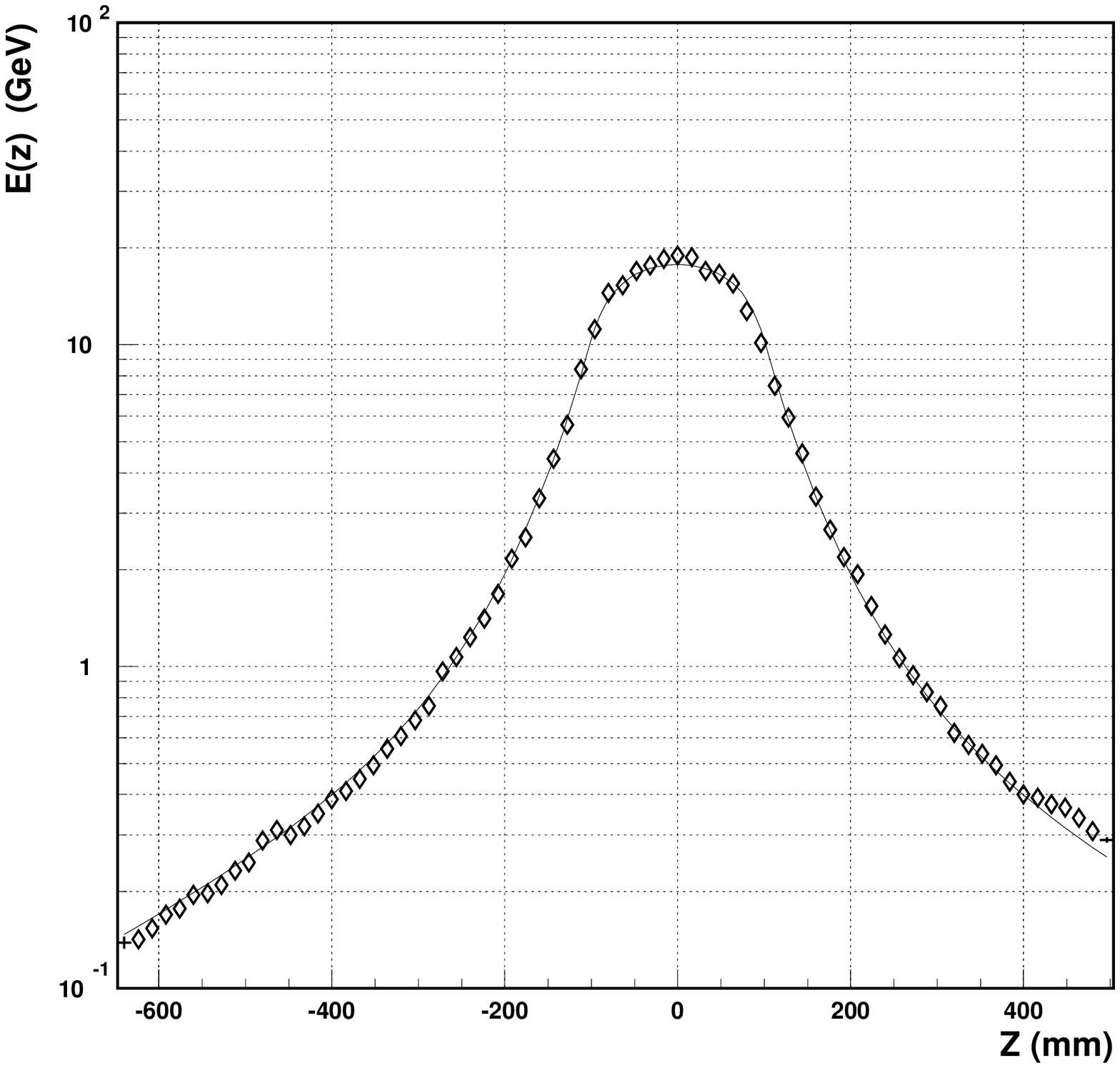,width=0.30\textheight}}
        &
        \mbox{\epsfig{figure=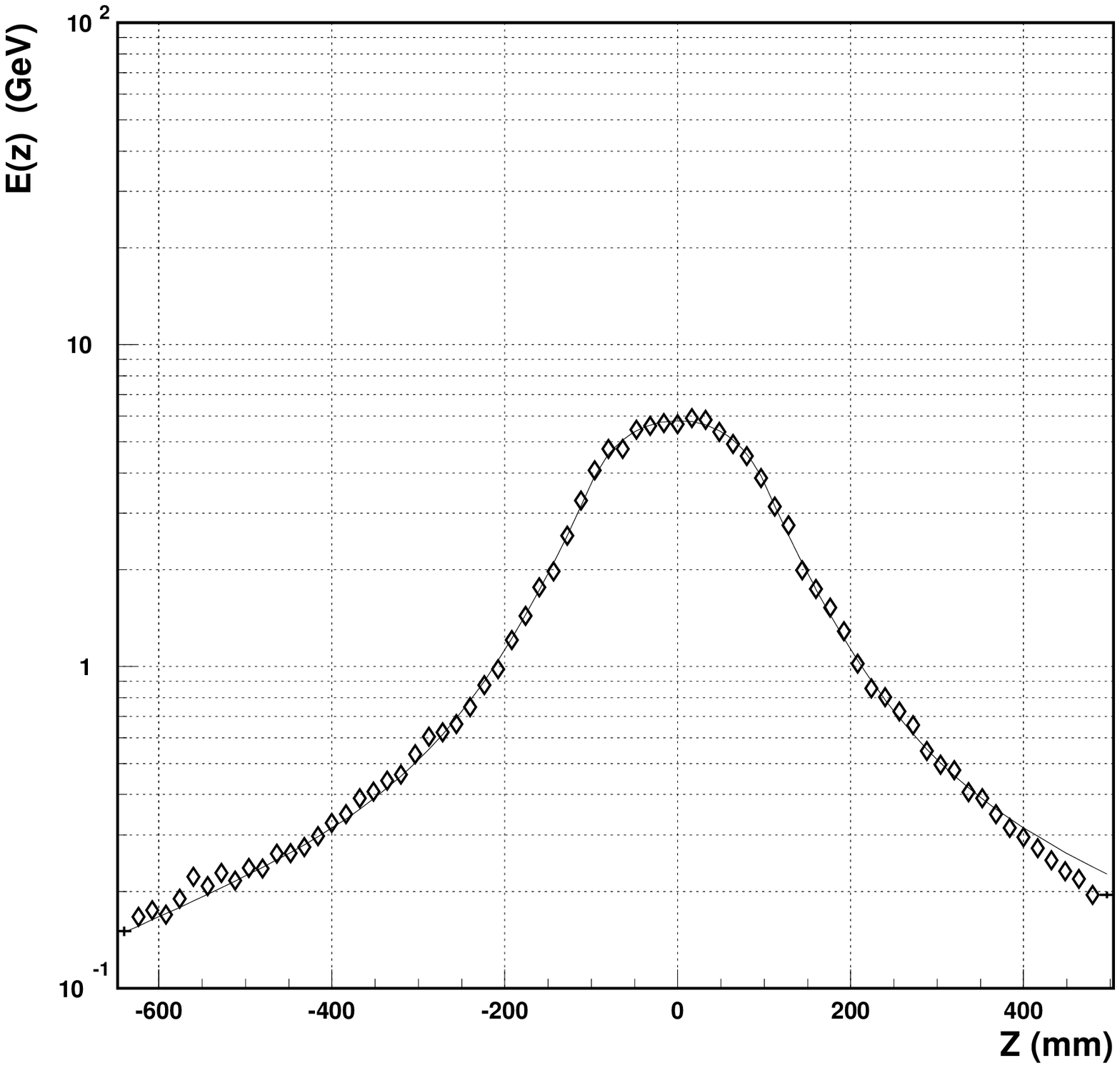,width=0.30\textheight}}
        \\
        \hline
        \end{tabular}
     \end{center}
      \caption{
        The energy depositions in towers of 1 $\div$ 4 depths as a 
        function of $Z$ coordinate. 
        Depth 1 --- up left,
        depth 2 --- up right,
        depth 3 --- down left,
        depth 4 --- down right.
        Curves are fits of equation (14) and (\ref{e023}) to the data.
       \label{fig:f5}}
\end{figure*}

\begin{figure*}[tbph]
     \begin{center}
        \begin{tabular}{|c|}
        \hline
        \mbox{\epsfig{figure=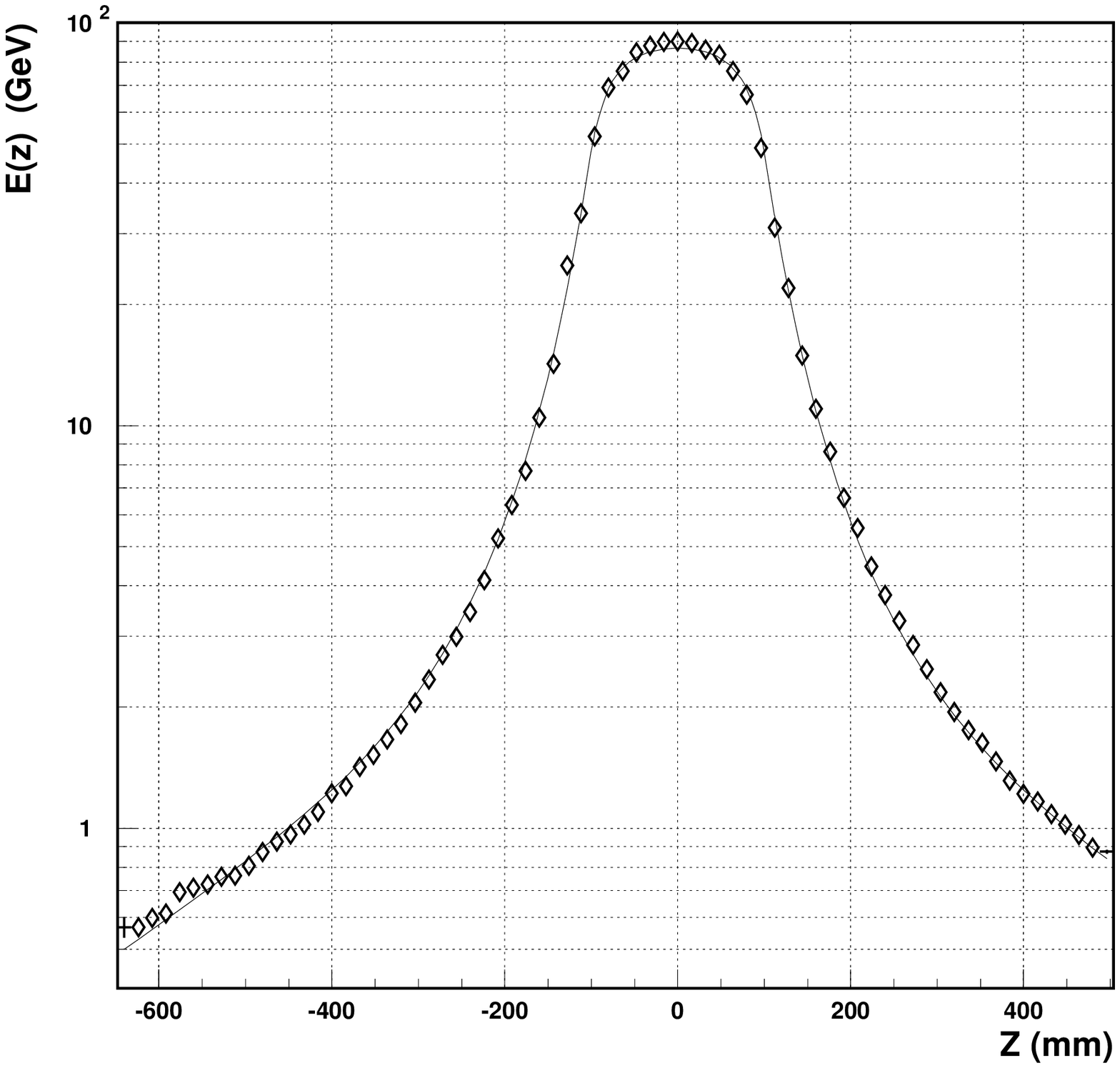,width=0.65\textheight}}
        \\
        \hline
        \end{tabular}
     \end{center}
      \caption{
        The energy depositions in towers summed overall calorimeter 
        depths as a function of $Z$ coordinate. 
        Curve is the result of fit by formula (14) and (\ref{e023}).
        Open circles are the Monte Carlo predictions 
        (GEANT-FLUKA+GHEISHA) given in  Table \ref{Tb003}.
       \label{fig:f6}}
\end{figure*}

Fig.~\ref{fig:f5} shows the energy depositions 
into towers  
for $1 \div 4$ depths as a function of $z$ coordinate of center of tower.
Fig.~\ref{fig:f6} shows the same for overall calorimeter
(the sum of histograms presented in Fig.~\ref{fig:f5}).
Due to the wide beam scan provided many different beam
impact locations with cells and using
information from all cells it is
succeeded to obtain detailed picture of transverse shower behaviour in
calorimeter.
The obtained spread of transverse shower dimensions is more than 1000 mm.
The energy depositions span a range of three orders of magnitude.
It can be seen the flat shoulder at $|z|$ coordinate less than 100 mm.
An immediate turnover occur as soon as $|z|$ reached the boundary of the cell.
Such picture of transverse shower behaviour was observed in other calorimeters
as well
\cite{acosta92},
\cite{gavrish91}.

We used these distributions in order to extract the underline 
mar\-gi\-nal density 
$f  (z)$.
By adding to statistical errors the electronic noise errors of 27 $MeV$/cell
\cite{cobal95},
the effective intercalibration error of $2 \%$
\cite{berger95}
and uncertainties of $4 \%$, arising from the nonzero entry angle of the
incident beam into the calorimeter, 
we obtained a good description of these distributions. 

The solid curves in Fig.~\ref{fig:f5} and  \ref{fig:f6}
are the results of fit with equation (\ref{e023}).
In compare with 
\cite{binon83},
\cite{gavrish91},
where the transverse profiles exists only for distances less than 
250 mm, the our more extended profiles (up to 650 mm) demand to introduce 
the third exponential.

The parameters $a_i$ and $\lambda_i$ obtained by fitting are listed in 
Table~\ref{Tb2}.
As can be seen from Table~\ref{Tb2} the slopes of exponentials,
$\lambda_i$, increase and the contribution of the first exponential, $a_1$,
decrease as shower develop.
The obtained values of $\lambda_1$ and $\lambda_2$ for overall 
calorimeter agree well with ones for conventional iron-scintillators 
calorimeter 
\cite{binon83} 
amount to $18\pm3$ mm and $57\pm4$ mm, respectively.

\begin{table}[tbph]
\caption{
        The parameters $a_i$ and $\lambda_i$ obtained by fitting 
        the transverse shower profiles for four depths and overall 
        calorimeter.
         \label{Tb2}}
\begin{center}
\begin{tabular}{@{}|c|c@{}|c@{}|c@{~}|c@{}|c@{}|c@{}|@{}}
\hline
        & $a_1$ 
        & $\lambda_1$, $mm$ 
        & $a_2$ 
        & $\lambda_2$, $mm$ 
        & $a_3$ 
        & $\lambda_3$, $mm$ 
\\
\hline
\hline
1
&$0.88\pm0.07$ 
&$17\pm2$ 
&$0.12\pm0.07$ 
&$48\pm14$ 
&$0.004\pm0.002$ 
&$430\pm240$
\\
\hline
2
&$0.79\pm0.06$ 
&$25\pm2$ 
&$0.20\pm0.06$ 
&$52\pm6$ 
&$0.014\pm0.006$ 
&$220\pm40$ 
\\
\hline
3
&$0.69\pm0.03$ 
&$32\pm8$ 
&$0.28\pm0.03$ 
&$71\pm13$ 
&$0.029\pm0.005$ 
&$280\pm30$ 
\\
\hline
4
&$0.41\pm0.05$ 
&$51\pm10$ 
&$0.52\pm0.06$ 
&$73\pm18$
&$0.07\pm0.03$ 
&$380\pm140$ 
\\ 
\hline
all
&$0.78\pm0.08$ 
&$23\pm1$ 
&$0.20\pm0.08$ 
&$58\pm4$
&$0.015\pm0.004$ 
&$290\pm40$ 
\\
\hline
\end{tabular}
\end{center}
\end{table}

The shower depth dependences of the parameters $a_i$ and $\lambda_i$ 
are displayed in Fig.~\ref{fig:5}.
As can be seen they demonstrated a linear behaviour.
The curves are fits of linear equations 
$a_i = \alpha_i + \beta_i \cdot x$ and  
$\lambda_i = \gamma_i + \delta_i \cdot x$.

\begin{figure*}[tbph]
     \begin{center}
        \begin{tabular}{|c||c|}
        \hline
        \mbox{\epsfig{figure=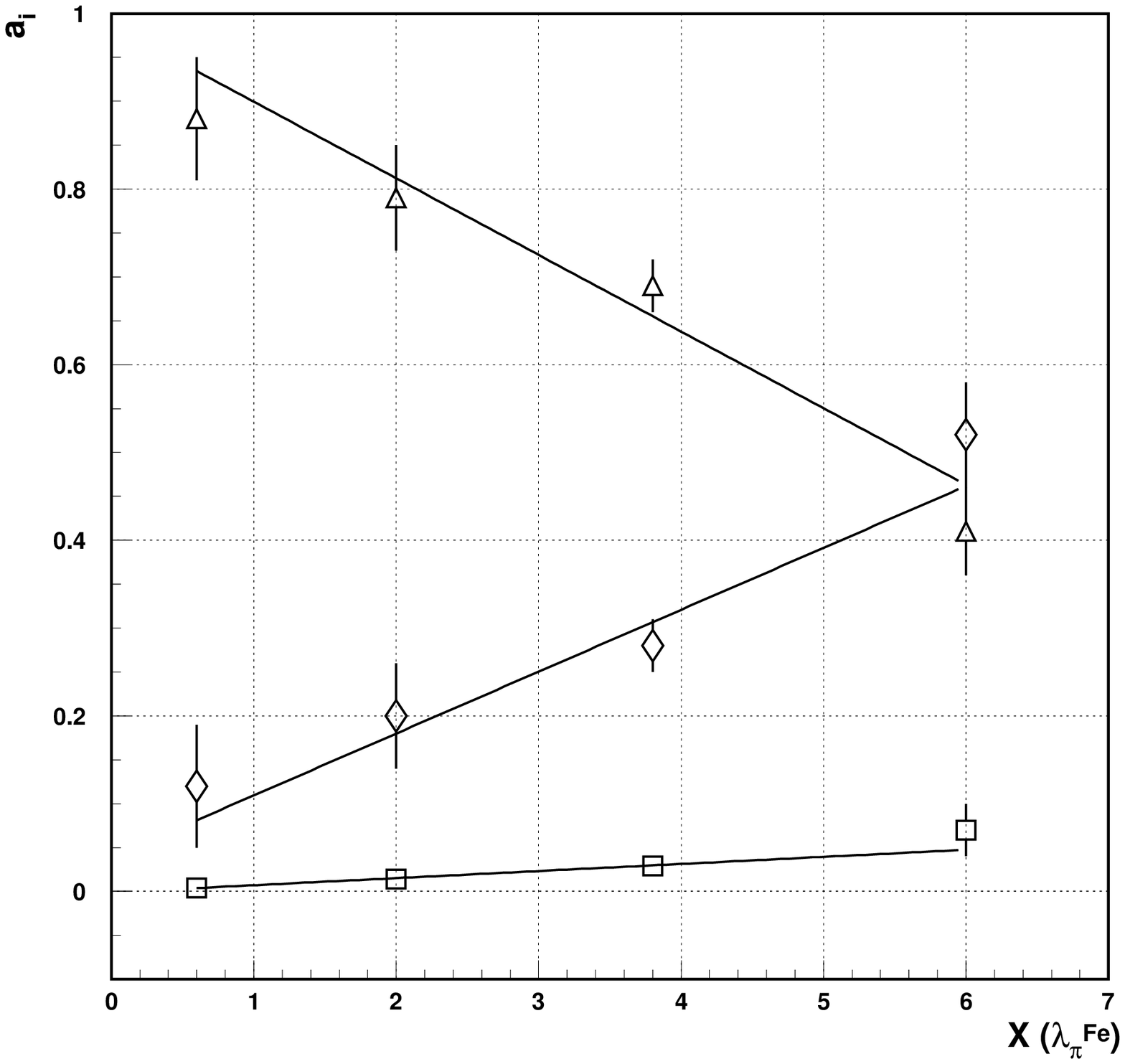,width=0.30\textheight}}
        &
        \mbox{\epsfig{figure=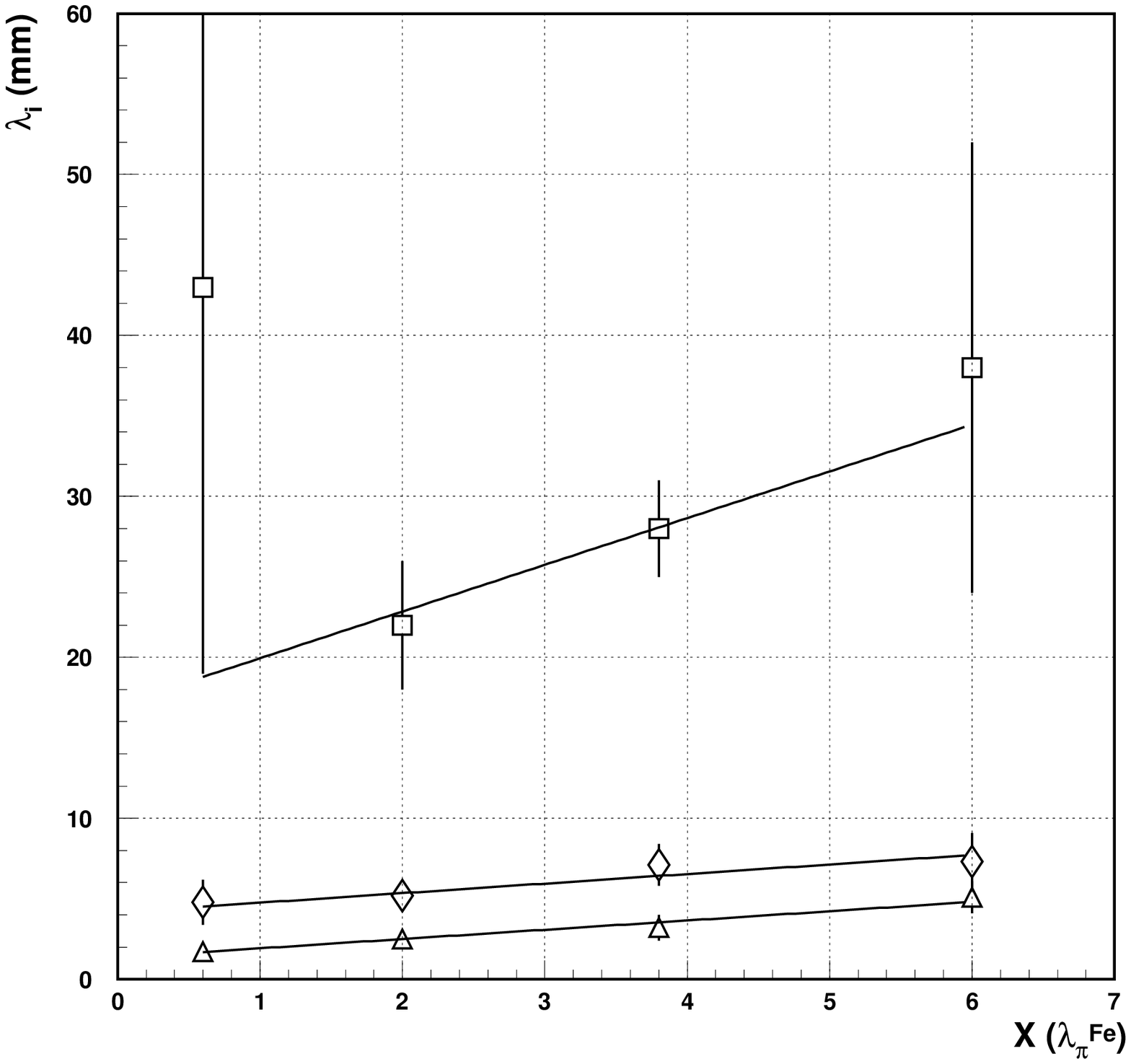,width=0.30\textheight}}
        \\
        \hline
        \end{tabular}
     \end{center}
      \caption{
        The shower depth dependences of the parameters $a_i$ and $\lambda_i$. 
       \label{fig:5}}
\end{figure*}

The values of parameters 
$\alpha_i$, $\beta_i$, $\gamma_i$ and $\delta_i$
are presented in the Table \ref{Tb02}.
It is interesting that in 
\cite{womersley88}
for the low-density-fine-grained flash chamber calorimeter
the linear behaviour of the slope exponential is also observed.
In the same time for uranium-scintillator $ZEUS$ calorimeter
some non-linear behaviour of slope of halo component measured 
at 100 $GeV$ has been demonstrated at interaction lengths more then
5 $\lambda$ in shower development
\cite{barreiro90}.

\begin{table}[tbph]
\caption{
         The values of parameters 
         $\alpha_i$, $\beta_i$, $\gamma_i$ and $\delta_i$.
         \label{Tb02}}
\begin{center}
\begin{tabular}{|c|c|c||c|c|c|}
\hline
        
        & $\alpha_i$ 
        & $\beta_i$, $1 / \lambda_{\pi}$
        & 
        & $\gamma_i$, $mm$ 
        & $\delta_i$, $mm / \lambda_{\pi}$ 
\\
\hline
\hline
        $a_1$
&       $0.99 \pm 0.06$ 
&       $- 0.089 \pm 0.015$ 
&       $\lambda_1$
&       $14 \pm 2$ 
&       $6 \pm 1$ 
\\
\hline
        $a_2$
&       $0.04 \pm 0.06$ 
&       $0.072 \pm 0.015$ 
&       $\lambda_2$
&       $42 \pm 10$ 
&       $6 \pm 4$ 
\\
\hline
        $a_3$
&       $- 0.001 \pm 0.002$ 
&       $0.008 \pm 0.002$ 
&       $\lambda_3$
&       $170 \pm 80$ 
&       $30 \pm 24$ 
\\
\hline
\end{tabular}
\end{center}
\end{table}

Table \ref{Tb003} presents the Monte-Carlo predictions 
for hadron shower energy depositions in the modules of TILECAL at 
centre coordinates of these modules obtained by using
the different hadronic simulation packages interfaced with GEANT
\cite{juste95}. 
The simulation energy cutoff for neutrons have been set to 1 $MeV$.
One can compare our data from Fig.\ \ref{fig:f6} and Monte-Carlo
predictions from Table \ref{Tb003} for the same $z$ coordinates.
Comparison shows worse agreement, with factor $> 3$, 
at distances of $\pm 400\ mm$ from shower kernel.

\begin{table}[tbph]
\caption{
  The average fractions of the energy deposited per module for 
  100 $GeV$ pions obtained by using the different hadronic simulation 
  packages interfaced with GEANT (G).
         \label{Tb003}}
\begin{center}
\begin{tabular}{|@{}l@{}|c|c|c|c|c@{}|}
\hline
\multicolumn{1}{|r|}{ Z, mm }
& $- 400$    & $- 200$  & $0$      & $200$    & $400$ \\ 
\hline
& \multicolumn{5}{|c@{}|}{E, \% } \\ 
\hline
\hline
{\small G-FLUKA+GHEISHA}
& $0.43\pm.02$ & $4.3\pm.1$ & $91\pm1$ & $4.4\pm.1$ & $0.43\pm.02$ \\
\hline
{\small G-FLUKA+MICAP}
& $0.20\pm.01$ & $3.2\pm.1$ & $94\pm1$ & $3.2\pm.1$ & $0.18\pm.01$ \\
\hline
{\small G-GHEISHA}
& $0.34\pm.01$ & $3.2\pm.1$ & $92\pm1$ & $4.5\pm.1$ & $0.38\pm.01$ \\
\hline
\end{tabular}
\end{center}
\end{table}

\subsubsection{Cumulative function}

Similar results were obtained from cumulative function distributions.
Cumulative function $F(z)$ was obtained as follows:
\begin{equation}
\label{eq-pr-2}
F (z) = \sum_{k=1}^{4} F^{k}(z),
\end{equation}
where 
$F^{k}(z)$ is the cumulative function for $k$-depth.
For each event $F^{k}(z)$ is
\begin{equation}
\label{eq-pr-3}
F^{k} (z) =  \sum_{i=1}^{i_{max}}
                        \sum_{j=1}^{5} E_{ijk},
\end{equation}
where 
$i_{max} = 1, \ldots, 5$ is the last number of tower in sum.


\begin{figure*}[tbph]
     \begin{center}
        \begin{tabular}{|c||c|}
        \hline
        \mbox{\epsfig{figure=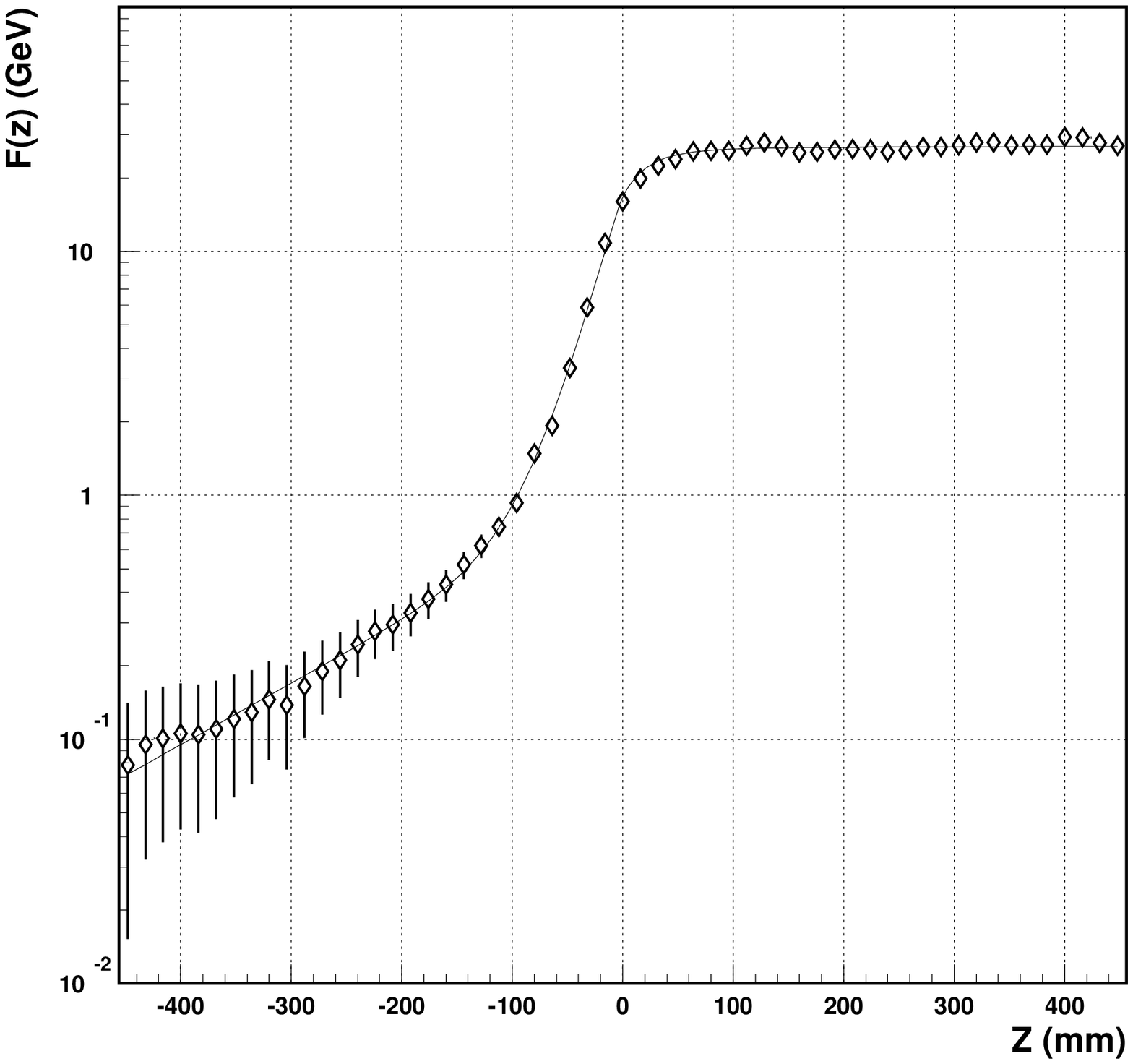,width=0.30\textheight}}
        &
        \mbox{\epsfig{figure=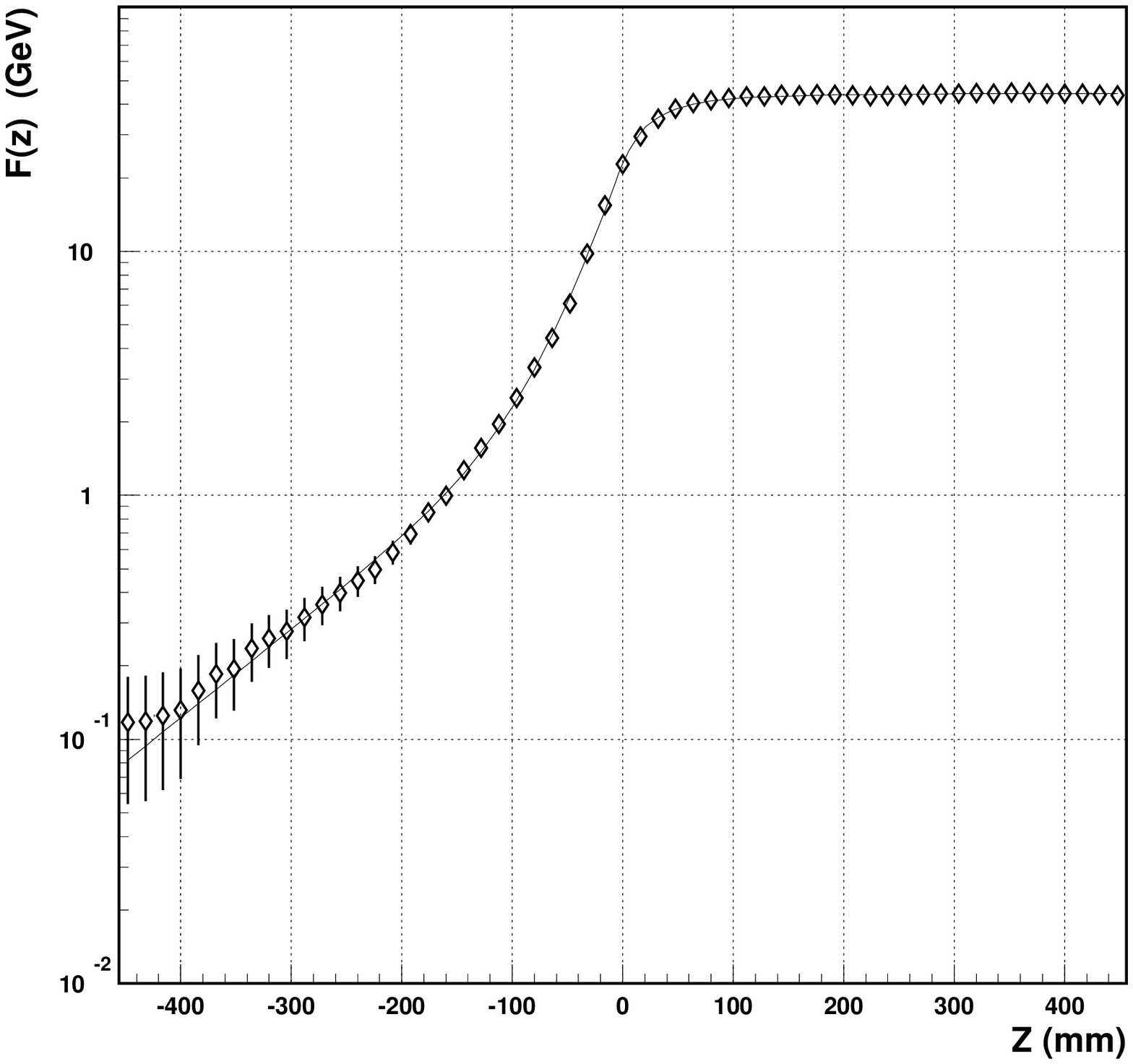,width=0.30\textheight}}
        \\
        \hline
        \hline
        \mbox{\epsfig{figure=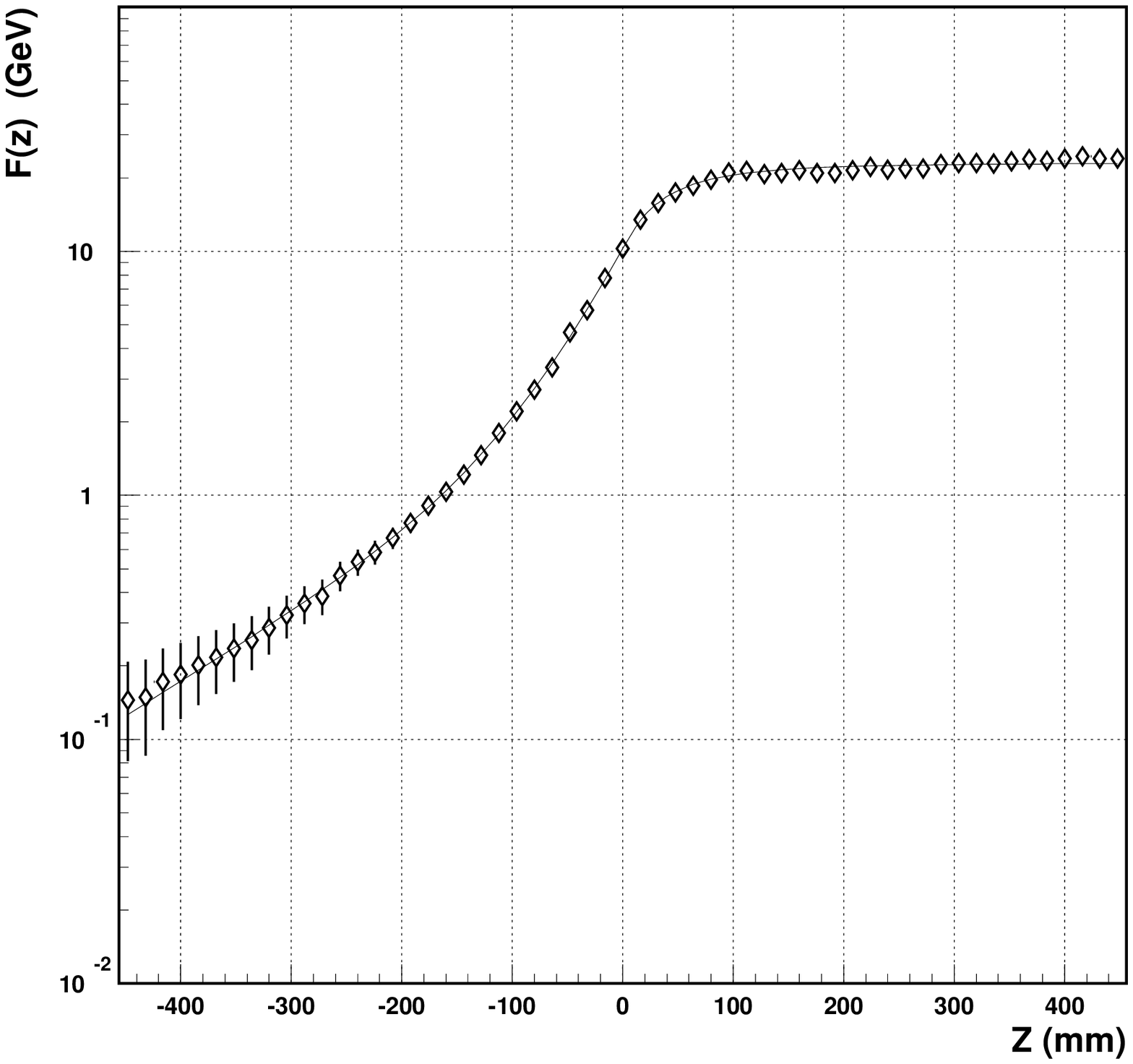,width=0.30\textheight}}
       &
        \mbox{\epsfig{figure=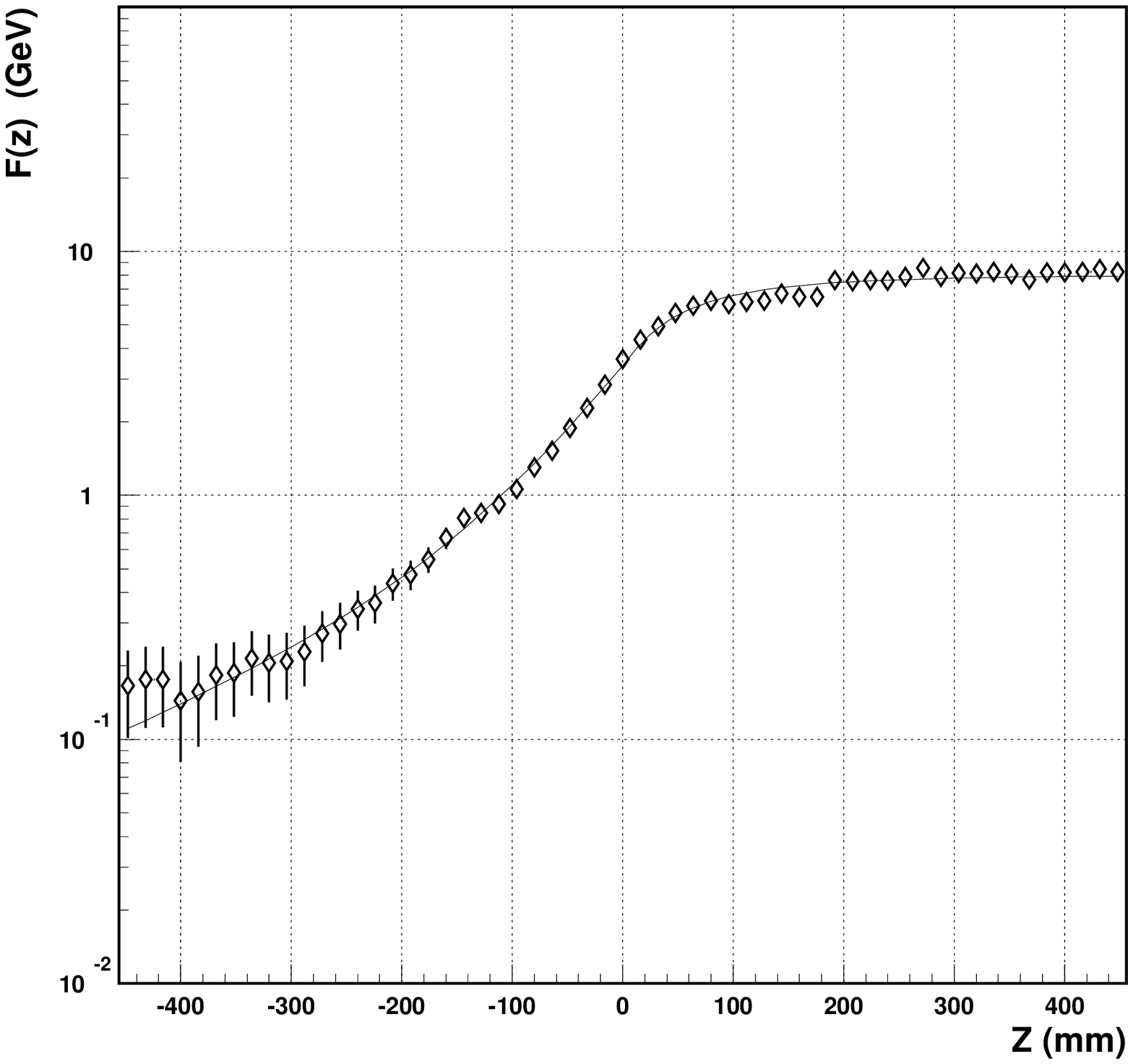,width=0.30\textheight}}
        \\
        \hline
        \end{tabular}
     \end{center}
      \caption{
        The cumulative functions $F (z)$ for four depths.
        Depth 1 --- up left,
        depth 2 --- up right,
        depth 3 --- down left,
        depth 4 --- down right.
        Curves are fits of equations (16) and (\ref{e23-2}) to the data.
       \label{fig:f7}}
\end{figure*}


\begin{figure*}[tbph]
     \begin{center}
        \begin{tabular}{|c||c|}
        \hline
        \mbox{\epsfig{figure=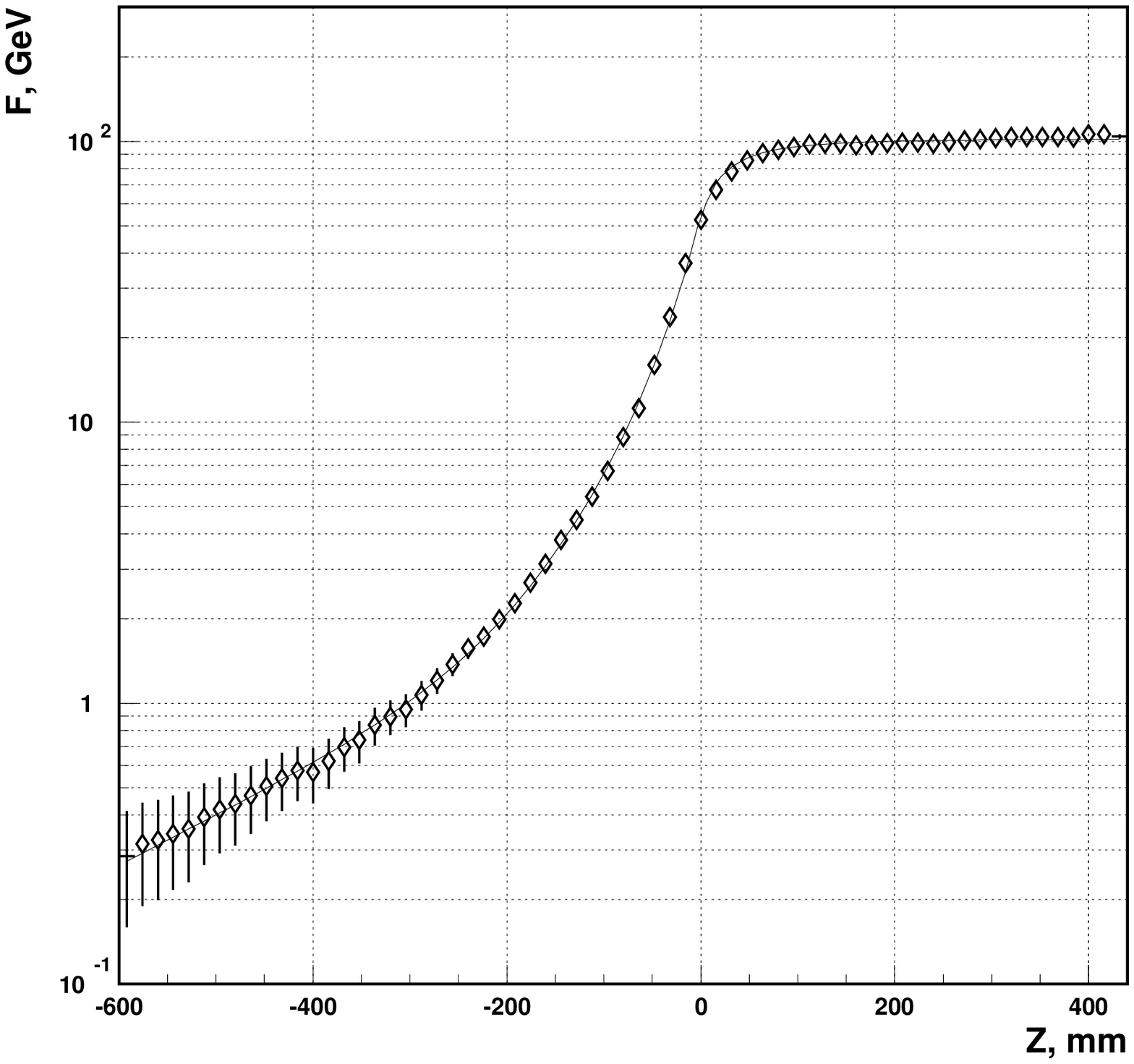,width=0.30\textheight}}
        &
        \mbox{\epsfig{figure=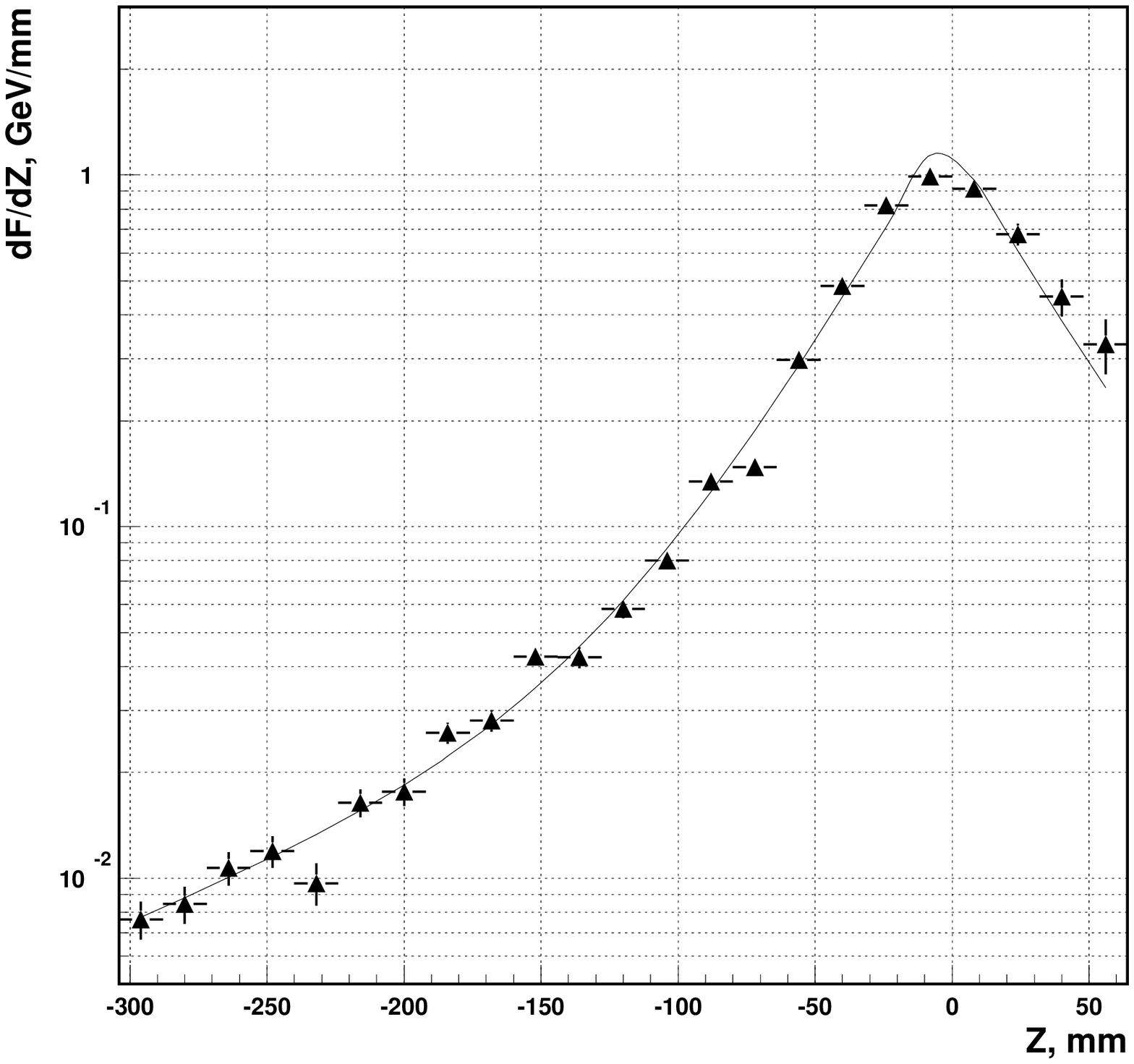,width=0.30\textheight}}
        \\
        \hline
        \end{tabular}
     \end{center}
      \caption{
        Left: The cumulative function $f  (z)$ for overall calorimeter.
        Curves are fits of equations (16) and (\ref{e23-2}) to the data.
        Right: The marginal density $f  (z)$ for overall calorimeter.
       \label{fig:f9}}
\end{figure*}

Fig.~\ref{fig:f7} and \ref{fig:f9} (left)
present the cumulative functions $F^{k}(z)$ for four depths
and for overall calorimeter.
The curves are fits of  equations (\ref{e23-2}) to the data.
The results of the cumulative function fits are less reliable 
and in the following we will use the results from energy depositions in tower.

The knowledge of the cumulative function allows us to 
determine directly according to (\ref{e30-01}) the marginal density 
$f  (z)$ without additional assumption about its form.
This is demonstrated in Fig.~\ref{fig:f9} (right) where the marginal density
$f  (z)$ extracted by the numerical differentiation $F (z)$ is shown.
Overlayed curve is a calculation of the expression (\ref{e21}) 
with parameters listed in Table \ref{Tb2}.
This curve well reproduces the such obtained marginal density.

Thus, the marginal densities  $f  (z)$ determined by three methods
(by using the energy deposition spectrum, the cumulative function
and the numerical differentiation of $F (z)$
are in reasonable agreement.

\subsubsection{Radial hadron shower energy density}

\begin{figure*}[tbph]
     \begin{center}
        \begin{tabular}{|c||c|}
        \hline
        \mbox{\epsfig{figure=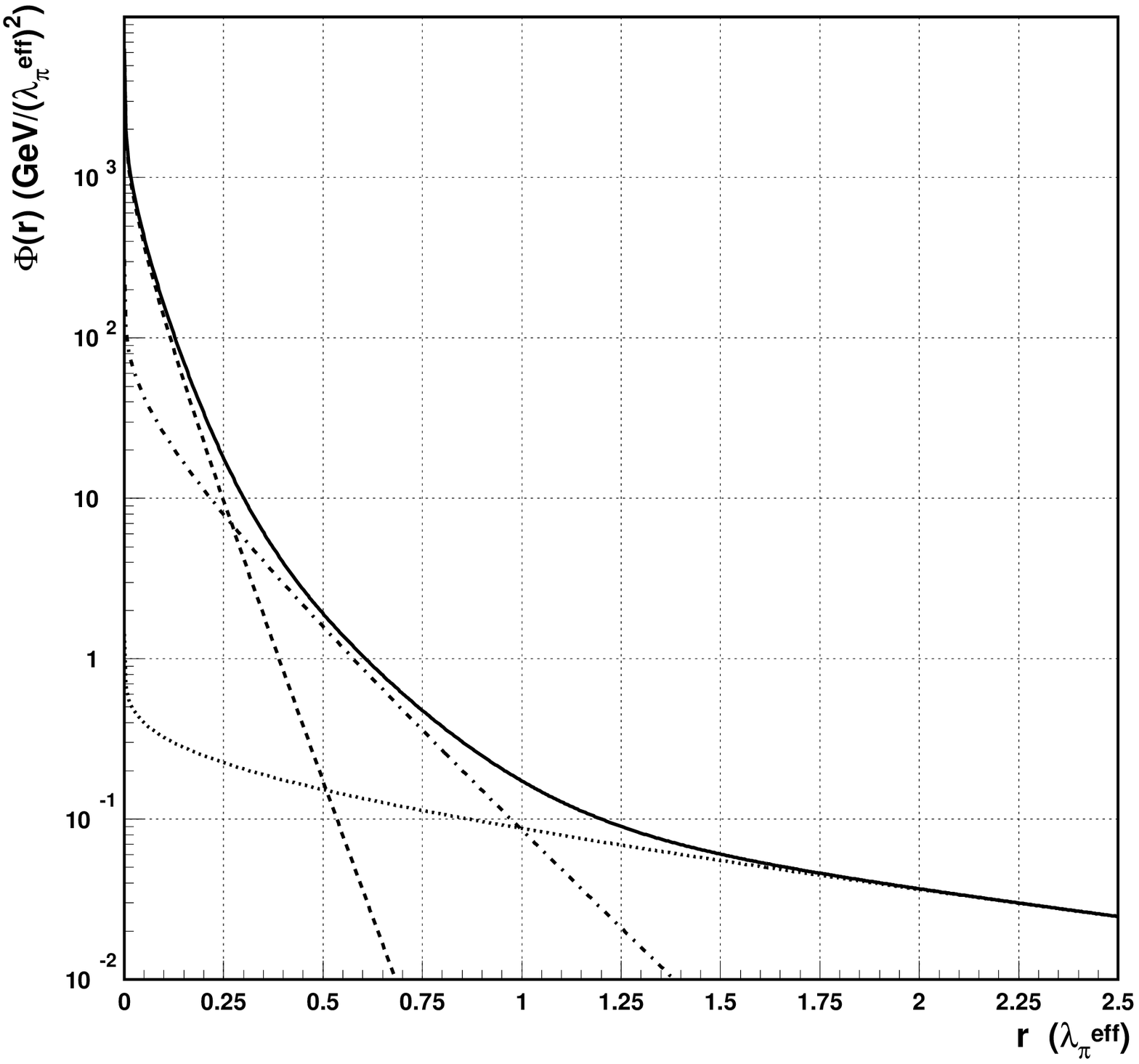,width=0.30\textheight}}
        &
        \mbox{\epsfig{figure=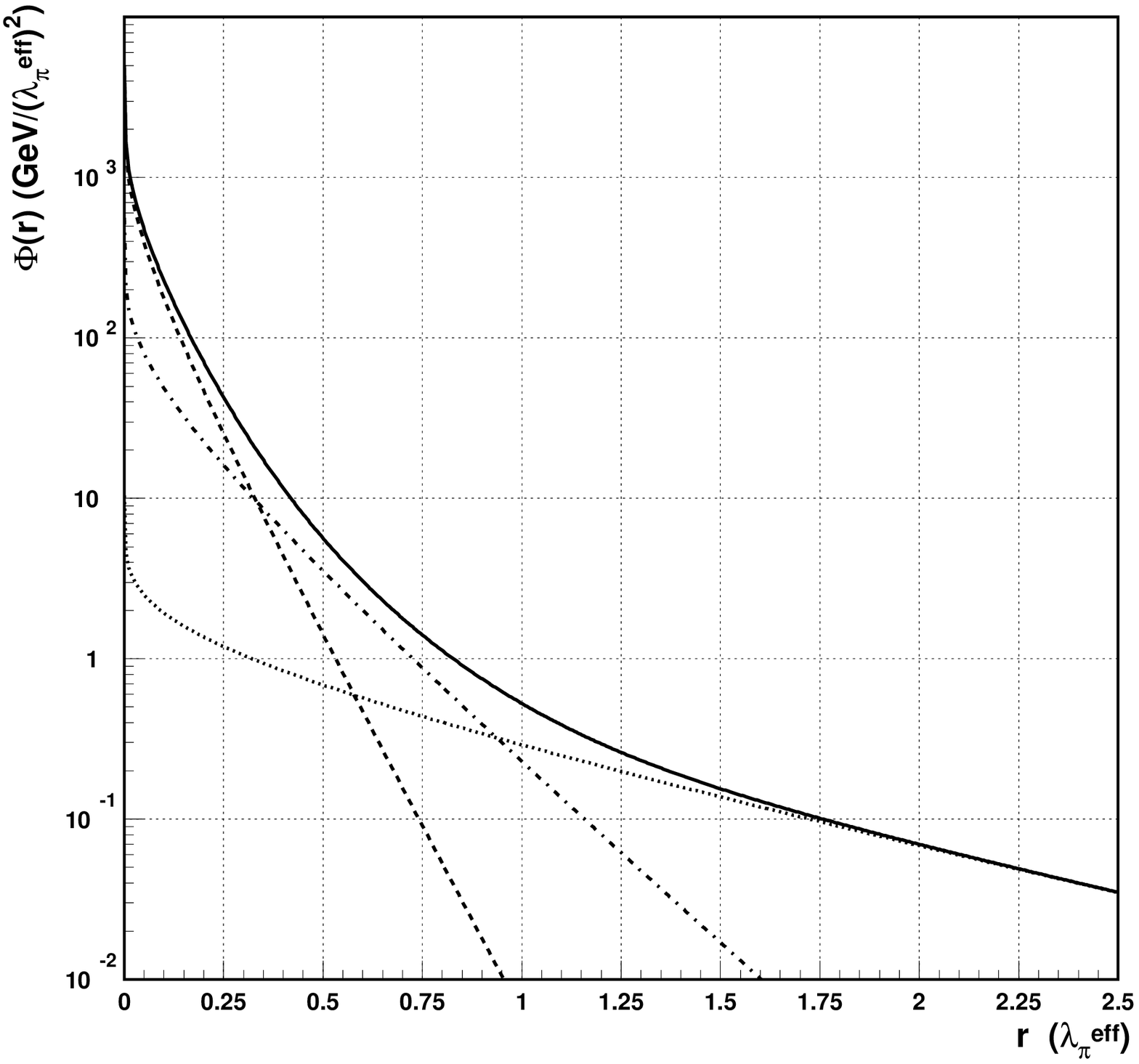,width=0.30\textheight}}
        \\
        \hline
        \hline
        \mbox{\epsfig{figure=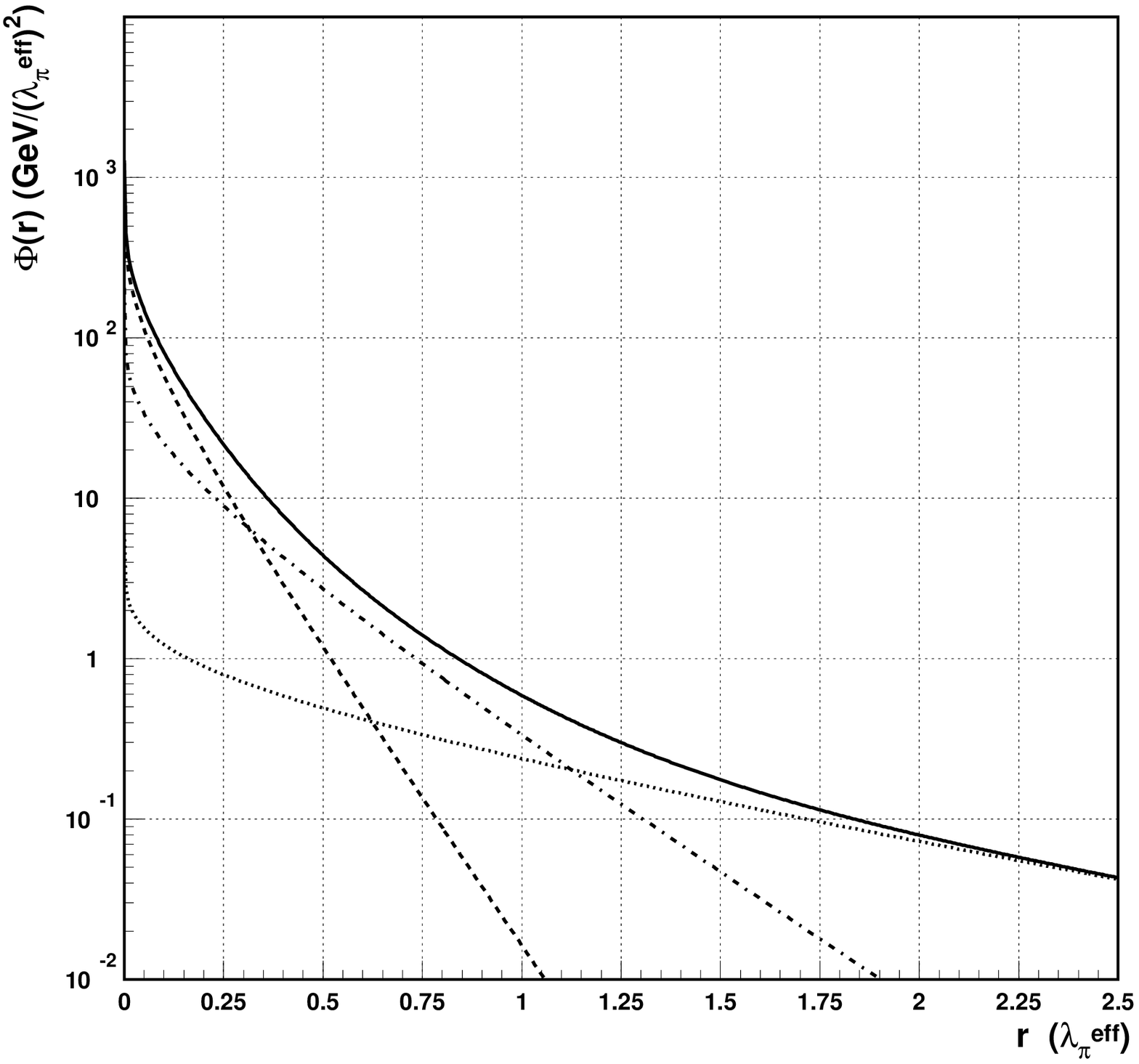,width=0.30\textheight}}
        &
        \mbox{\epsfig{figure=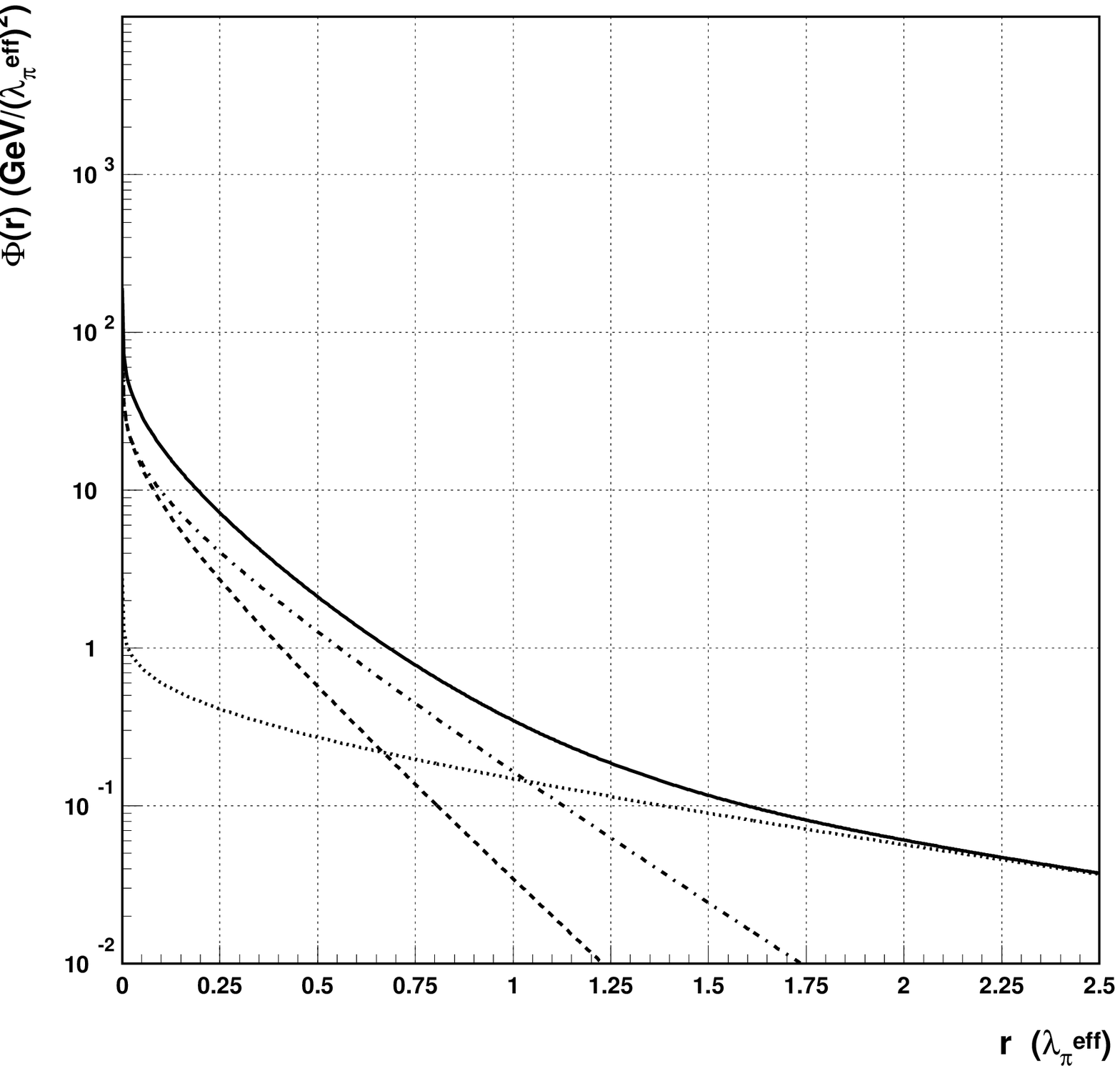,width=0.30\textheight}}
        \\
        \hline
        \end{tabular}
     \end{center}
      \caption{
        The radial 
        energy density, $\Phi (r)$, as a function of $r$ for TILECAL
        for various depths:
        solid lines --- the energy densities $\Phi (r)$,
        dashed lines --- the contribution from the first term,
        dash-dotted lines --- the contribution from the second term,
        dotted lines --- the contribution from the third term.
        Depth 1 --- up left,
        depth 2 --- up right,
        depth 3 --- down left,
        depth 4 --- down right.
       \label{fig:f17}}
\end{figure*}

\begin{figure*}[tbph]
     \begin{center}
        \begin{tabular}{|c||c|}
        \hline
      \mbox{\epsfig{figure=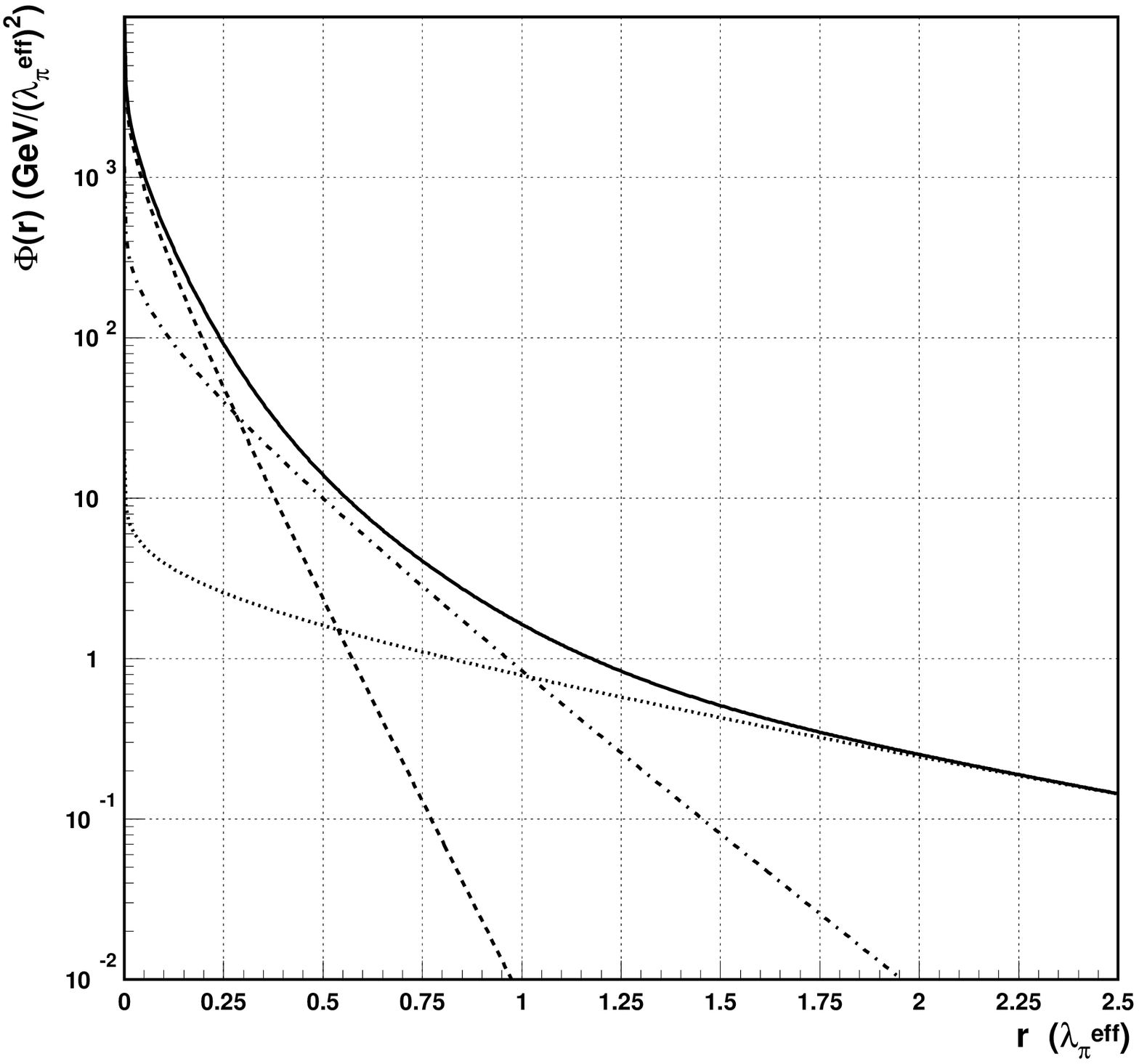,width=0.30\textheight}}
        &
      \mbox{\epsfig{figure=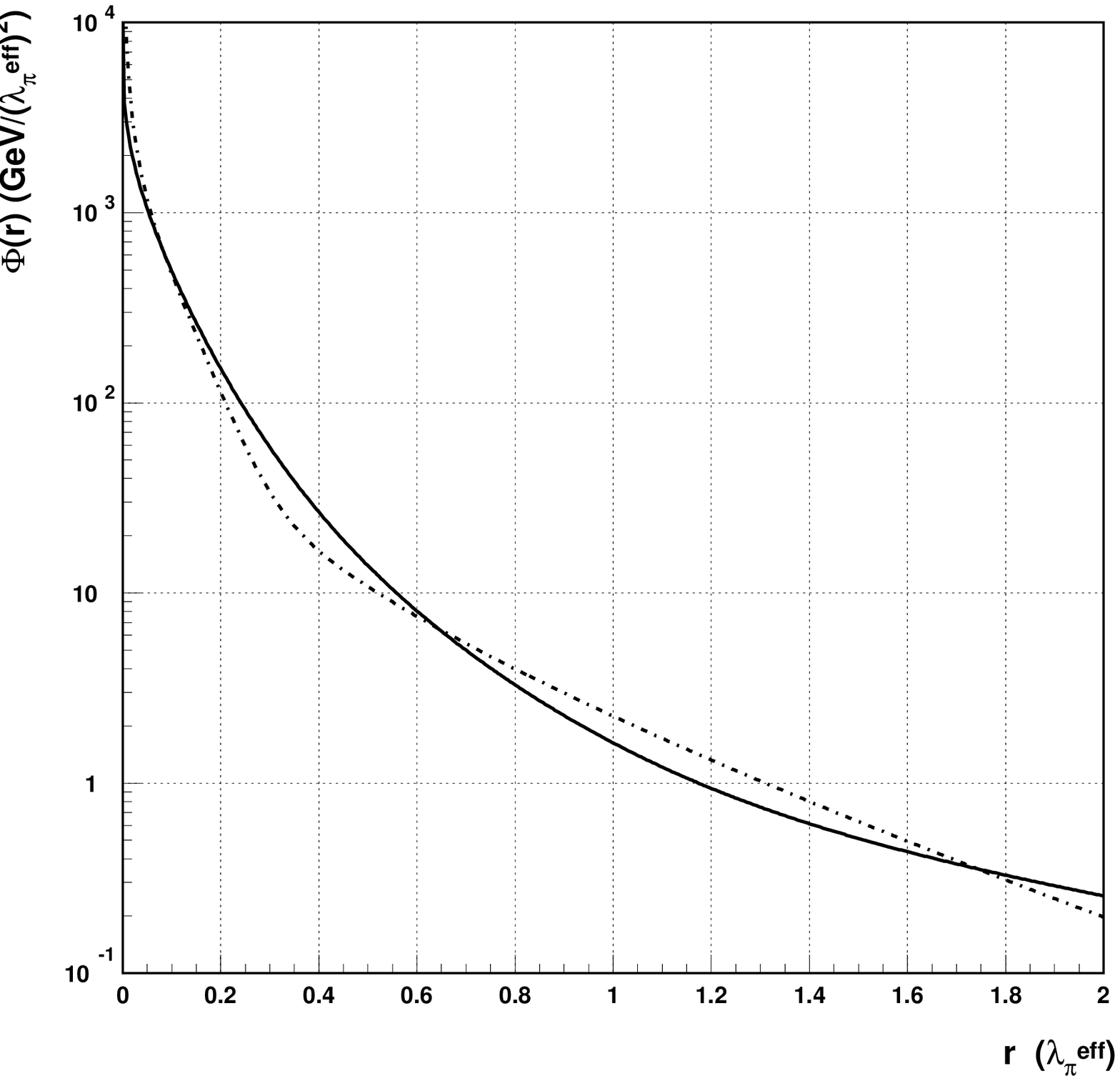,width=0.30\textheight}}
        \\
        \hline
        \end{tabular}
     \end{center}
       \caption{
        Left: The radial energy density as a function of $r$ 
        (in units of $\lambda_f$) for TILECAL (solid line),
        the contribution to $\Phi (r)$ 
        from the first term (dashed line),
        the contribution to $\Phi (r)$ 
        from the second term (dash-dotted  line),
        the contribution to $\Phi (r)$ 
        from the third term (dotted  line).
        Right: The comparison of the radial energy densities
        as a function of $r$ (in units of $\lambda_f$) for
        TILECAL (solid line) and SPACAL (dash-dotted line).
       \label{fig:f10}}
\end{figure*}

Fig.\ \ref{fig:f17} and Fig.\ \ref{fig:f10} (left)
show the radial shower energy density functions, $\Phi (r)$, calculated 
by formula (\ref{e23}) with parameters from table \ref{Tb2}.
The contributions of different terms are also shown.
The data of SPACAL calorimeter calculated by formula (\ref{e1})
with 
${\lambda}_1 = 140\ mm$
and
${\lambda}_2 = 42.4\ mm$
are shown for comparison on Fig.~\ref{fig:f10} (right).
Since in this work the lateral profiles are presented
in term of the measured picocoulombs the
density $\Phi (r)$
was transformed into $GeV$ energy scale by using the energy deposit constant
of 4 $pC/GeV$ \cite{acosta92}.

As can be seen at general reasonable agreement the curve of SPACAL lies 
systematically below the TILECAL one beyond $1.5\ \lambda_{\pi}^{eff}$.

\subsubsection{Radial containment}

\begin{figure*}[tbph]
     \begin{center}
        \begin{tabular}{|c|c|}
        \hline
      \mbox{\epsfig{figure=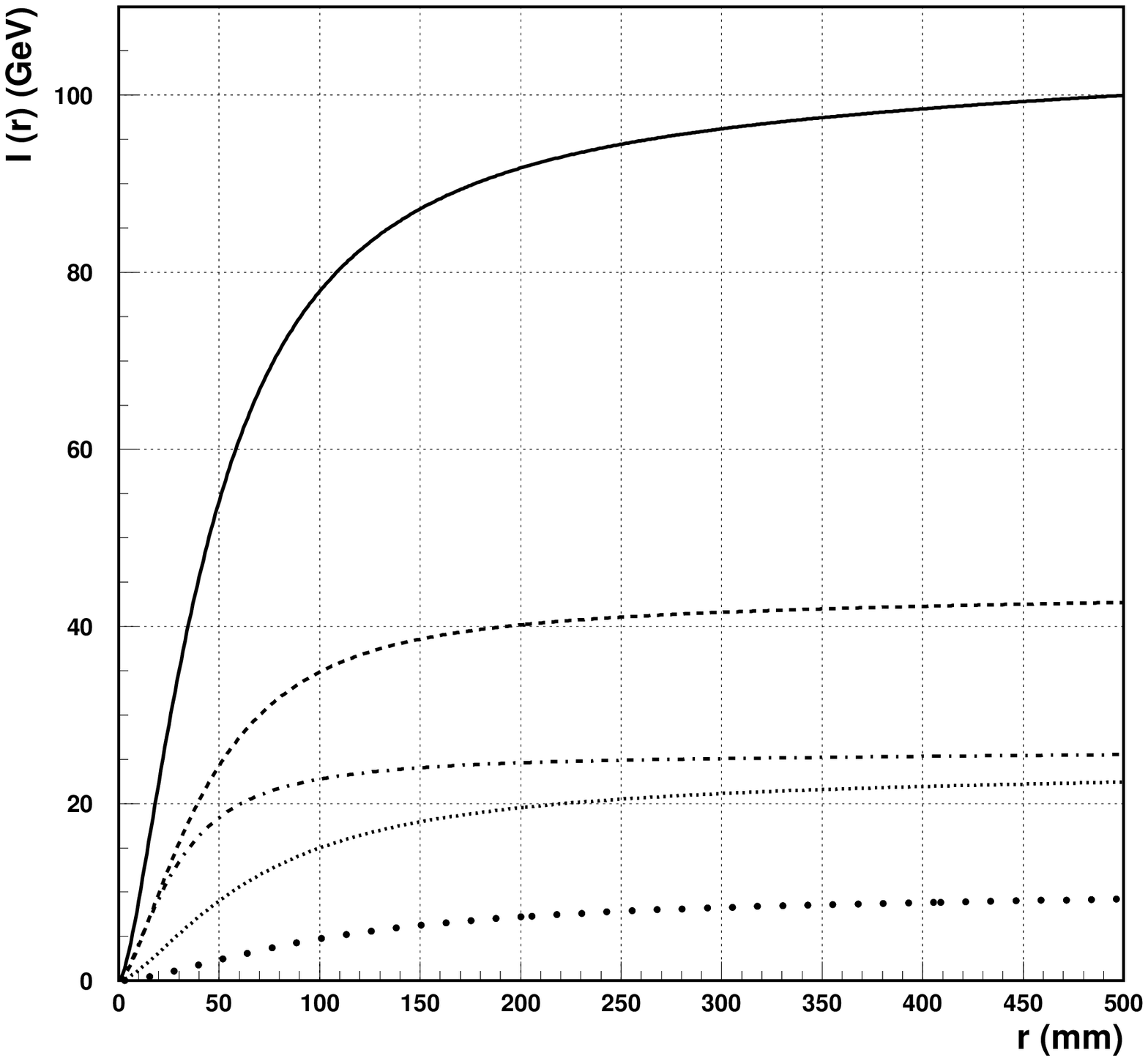,width=0.30\textheight}}
        &
        \mbox{\epsfig{figure=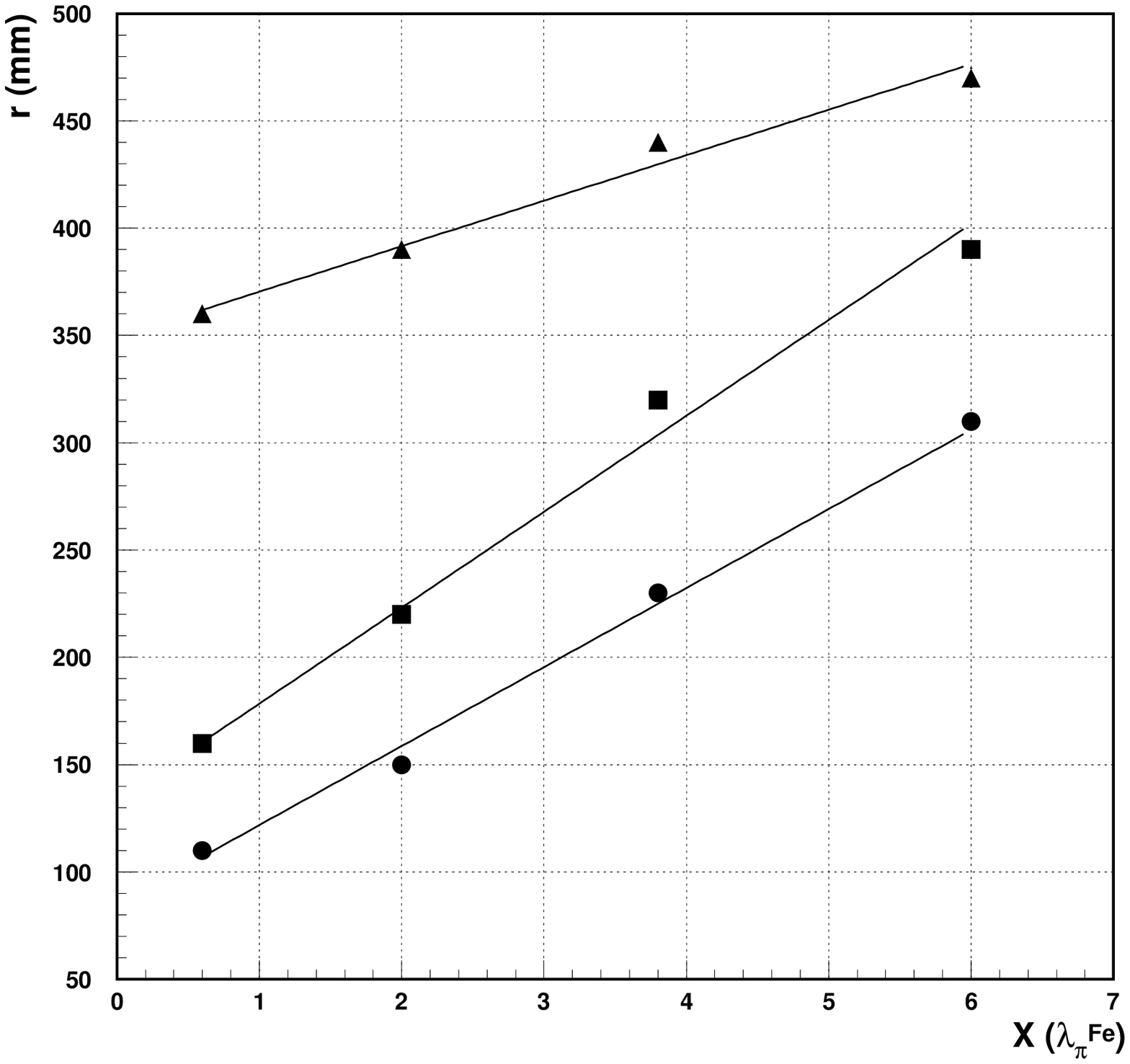,width=0.30\textheight}}
        \\
        \hline
        \end{tabular}
     \end{center}
       \caption{
        Left:
        The containment of shower $I (r)$ (solid line)
        as a function of radius for overall TILECAL calorimeter.
        Dash-dotted line --- the contribution from first depth,
        dashed line --- the contribution from second depth,
        thick dotted line --- the contribution from third depth,
        thin dotted line --- the contribution from fourth depth.
        Right:
        The radii of contained cylinders for given shower containment
        as a function of depths: 
        black circle -- $90 \%$ of containment, 
        black square -- $95 \%$,
        black triangle -- $99 \%$.
        The curves are drawn to guide the eye.
       \label{fig:f11}}
\end{figure*}



One of the important questions concerns the shower transverse
dimensions and its longitudinal development. 

The parameterisation of the radial density function, $\Phi (r)$,
have been integrated to yield the shower containment as a function 
of radius, $I (r)$.
Fig.\ \ref{fig:f11} (left)
shows the transverse containment of the pion shower,
$I (r)$, as a function of $r$ for four depths and overall calorimeter.

\begin{table}[tbph]
\caption{
        The radii of contained cylinders for given shower containment
        for TILECAL (for various depths and overall) and SPACAL (overall)
calorimeters.
        \label{Tb3}}
\begin{center}
\begin{tabular}{|l|c|c|c|c|}
\hline
Calorimeter & x in $\lambda_{\pi}^{Fe}$ 
& \multicolumn{3}{|c|}{$r$, in $\lambda_{\pi}^{eff}$ }\\
\cline{3-5}
            &           & 90 \% & 95 \% & 99 \% \\
\hline
 TILECAL    & 0.59      & 0.44    & 0.64    & 1.43   \\
\cline{2-5}
            & 1.97      & 0.60    & 0.88    & 1.55   \\
\cline{2-5}
            & 3.74      & 0.92    & 1.27    & 1.75   \\
\cline{2-5}
            & 5.92      & 1.24    & 1.55    & 1.87   \\
\cline{2-5}
            & overall   & 0.72    & 1.04    & 1.67   \\
\hline
SPACAL      & overall   & 0.86    & 1.19    & 1.72   \\
\hline
\end{tabular}
\end{center}
\end{table}



In Table \ref{Tb3} and Fig.\ \ref{fig:f11} (right) 
the radii of cylinders for given shower containment
(90\%, 95\%, 99\%) extracted from Fig.\ \ref{fig:f11} (left)
as a function of depth are shown.
Solid lines are the linear fit to the data.
As can be seen these containment  radii increase linearly with the depth.
The linear increase of 95
depth is also observed for Fe-scintillator calorimeter at 50 and 140
$GeV$ 
\cite{holder78}.
For overall TILECAL calorimeter the 99\% containment radius is equal to 
$1.7 \pm 0.1$ $\lambda_{\pi}^{eff}$.
In the last row of this Table
the corresponding data for SPACAL calorimeter for
80 $GeV$ $\pi$-mesons
are given (a pions grid scan at an angle of $2^o$ with respect to
fiber direction).
Note that in the case of Pb-scintillator calorimeter SPACAL,
having the same $\lambda_{\pi}^{eff}$ 
(see Table \ref{tb-a22} in Appendix 2),
the shower containment radii are similar to obtained for iron-scintillator 
calorimeter TILECAL.   

It is interesting to note that it is mistaken to consider 
as the measure of the transverse shower containment the one obtained from
the marginal density $f  (z)$ or the energy depositions in strips as
have been made in 
\cite{womersley88}.
In this work for $99 \%$ containment have been obtained the value of 
$1.2\ \lambda_I$ 
and the conclusion have been made that 
``their result is consistent with the ``rule of thumb'' 
that a shower is contained within a cylinder of radius equal 
to the interaction length of a calorimeter material
\cite{amaldi80}''.
But it is showed by our and SPACAL measurements that the value of radius
of 99\% contained shower cylinder amounts to about two interaction length.
Reduced value of this radius obtained from $f  (z)$ is due to that 
according to (\ref{e4}) represents the integrated function 
$\Phi (r)$.
In our case if we will use as a measure of shower containment the 
half-width of integrated $f  (z)$ estimated from $F (z)$ 
in Fig.\ \ref{fig:f9} (left) we obtain the value of $300\ mm$ or 
$1.2\ \lambda_{\pi}^{eff}$ that agrees with 
\cite{womersley88}.

\subsection{Longitudinal Profile}

Here we are concerned with the differential deposition of energy
$ \Delta E/ \Delta x$
as a function of $x$, the distance along the shower axis.
In Table \ref{long-1} the average energy shower depositions in various depths,
$E_o$, the normalised to one interaction length $\lambda_{\pi}^{Fe}$
energy depositions,
$\Delta E / \Delta x$, the lengths of depths, $L$, the effective iron
lengths of depths,
$L_{eff}$, the lengths of depths in units of $\lambda_{\pi}^{Fe}$,
the centers of depth intervals in units of $\lambda_{\pi}^{Fe}$, 
$x$, are given.
In these calculations the value of $\lambda_{\pi}^{Fe} = 207\ mm$ 
has been used (see Appendix 3).

\begin{table}[tbph]
\caption{
        Average energy shower depositions in various depths.
        \label{long-1}}
\begin{center}
\begin{tabular}{|c|c|c|c|c|c|}
\hline
 $x$ in $\lambda_{\pi}^{Fe}$
&$\Delta x$ in $\lambda_{\pi}^{Fe}$
&$L$, mm
&$L_{Fe}$, mm
&$E_o$, $GeV$
&$\Delta E / \Delta x$, $GeV$\\
\hline
\hline
0.59  & 1.18  & 300  & 245  & 25.5$\pm$0.3  & 21.6$\pm$0.3\\
\hline
1.97  & 1.58  & 400  & 327  & 43.6$\pm$0.2  & 27.6$\pm$0.1\\
\hline
3.74  & 1.97  & 500  & 408  & 22.4$\pm$0.1  & 11.2$\pm$0.1\\
\hline
5.92  & 2.37  & 600  & 490  &  8.5$\pm$0.5  &  3.6$\pm$0.2\\
\hline
\end{tabular}
\end{center}
\end{table}

In Fig.~\ref{fig:f15} the our quantities  $\Delta E / \Delta x$ 
(open circle)
together with the data of 
\cite{huges90} (open triangles)
and Monte Carlo predictions ($GEANT-FLUKA + MICAP$, diamonds)
\cite{juste95}
are shown.
The agreement is observed.
So, as to longitudinal energy deposition our calorimeter  with longitudinal
orientation of the scintillating tiles agrees with the one for a conventional
iron-scintillator calorimeters.

\begin{figure*}[tbph]
     \begin{center}
        \begin{tabular}{|c|c|}
        \hline
        \mbox{\epsfig{figure=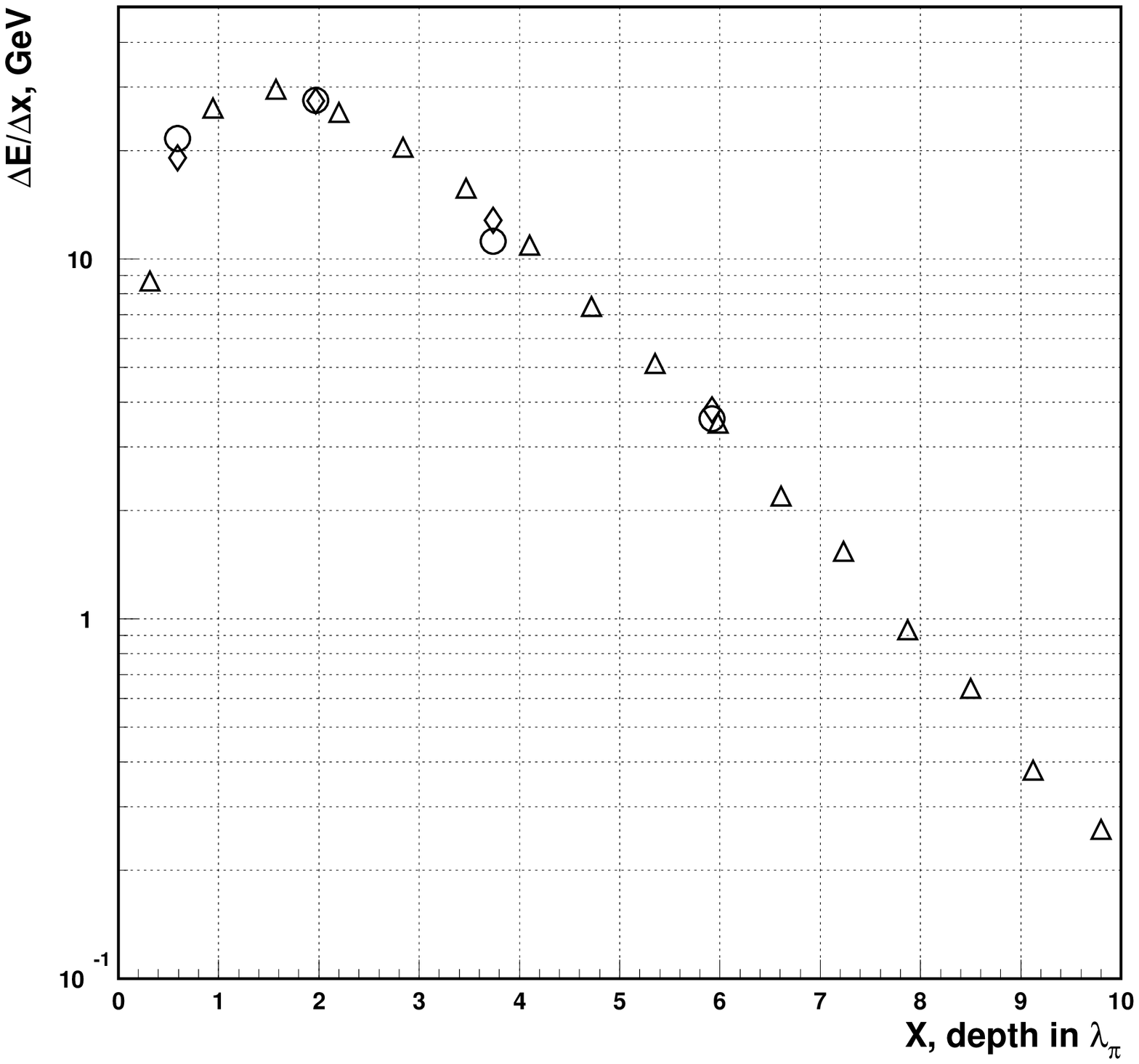,width=0.30\textheight}} 
        &
        \mbox{\epsfig{figure=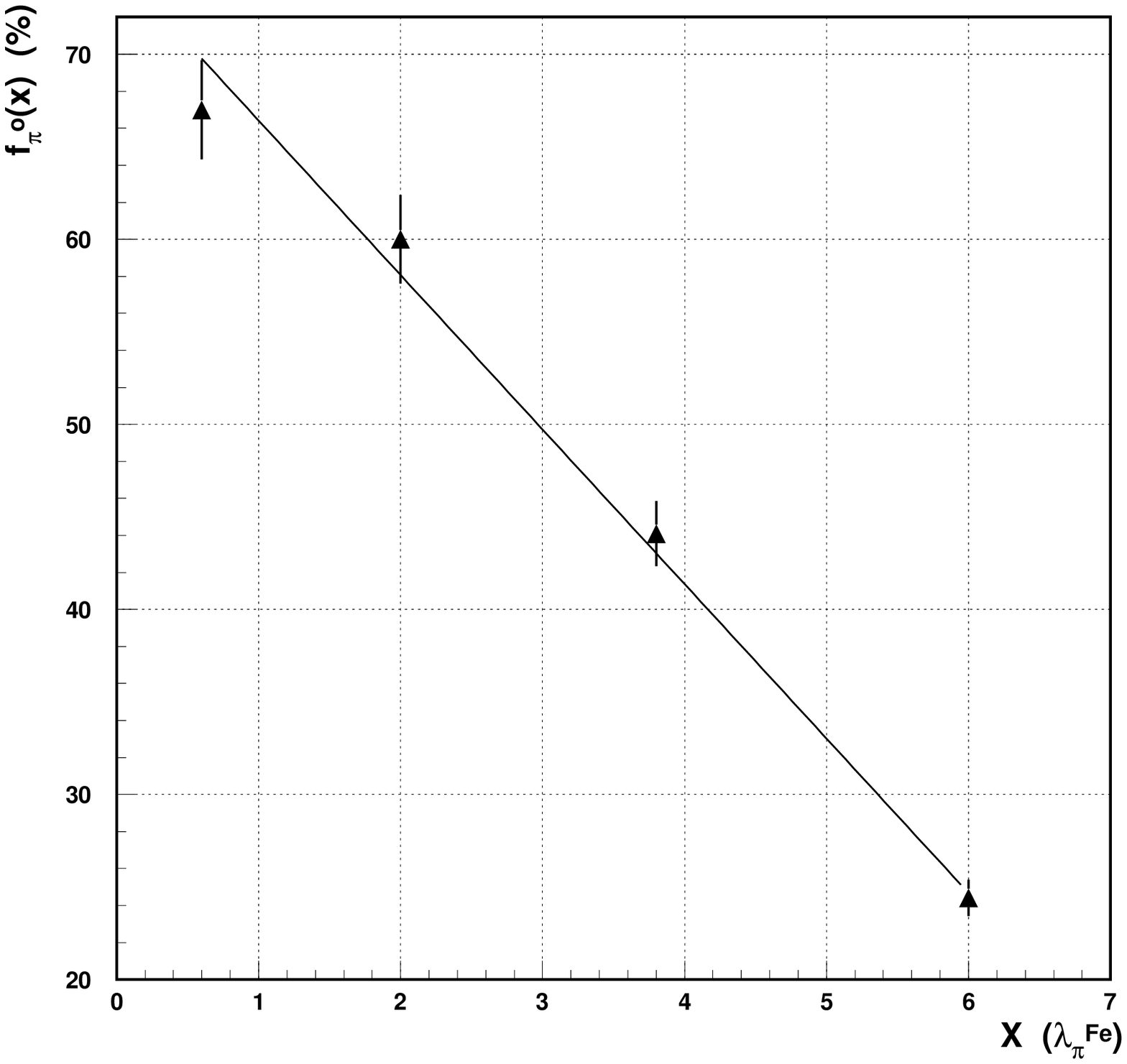,width=0.30\textheight}}
        \\
        \hline
        \end{tabular}
     \end{center}
       \caption{
        Left: 
        Longitudinal profile as a .function of 
        longitudinal coordinate $x$ in units 
        $\lambda_{\pi}^{Fe}$. 
        Open circles are our data,
        open triangles are conventional calorimeter data, 
        diamonds are the Monte Carlo predictions.
        Right:
        The fractions of the ``electromagnetic'' parts of showers 
        in various depths as a function 
        of $x$.
       \label{fig:f15}}
\end{figure*}

\subsection{``Electromagnetic'' fraction of a hadronic shower}

The lateral shower profile information may be used for determination of the 
average fraction of energy going 
into $\pi^{o}$ production in a hadronic shower,
$f_{\pi^{o}}$
\cite{acosta92}.
Following 
\cite{acosta92}
we assume that the ``electromagnetic'' part of hadronic shower is the 
prominent central core, i.e.\ in our case the first term.
The integrated contribution from this term is 
$f_{\pi^{o}} = (55 \pm 3) \%$.
This value may be compared with 
the one of $56 \%$ as simulated by $FLUKA$ for our calorimeter
\cite{juste95},
the one of $59 \%$ as simulated by $CALOR$
for iron calorimeter 
\cite{gabrial94}
and the one of $54 \%$ as obtained by using the lead-scintillating fiber 
$SPACAL$ calorimeter
\cite{acosta92}.

The observed $\pi^{o}$ fraction, $f_{\pi^{o}}$, is related to the 
intrinsic actual fraction, $f^{\prime}_{\pi^{o}}$, by the relation
\cite{groom90}, 
\cite{wigmans88}:
\begin{equation}
f_{\pi^{o}} = 
\frac{ e/h \cdot f^{\prime}_{\pi^{o}}}{(e/h - 1) 
\cdot f^{\prime}_{\pi^{o}} + 1}.
\label{e506-1}
\end{equation}
There are two analytic forms for the intrinsic $\pi^{o}$
fraction suggested by Groom
\cite{groom90}
\begin{equation}
f^{\prime}_{\pi^{o}} = 1 - {(\frac{E}{E_o^{\prime}})}^{(m-1)}
\label{e506-2}
\end{equation}
and Wigmans \cite{wigmans88}
\begin{equation}
f^{\prime}_{\pi^{o}} = k \cdot ln ({\frac{E}{E_o^{\prime}}}),
\label{e506-3}
\end{equation}
where $E_o^{\prime} = 1\ GeV$, $m = 0.85$, $k = 0.11$.
We calculated the values of $f_{\pi^{o}}$ using a value of $e/h = 1.23$ for 
our calorimeter at $\Theta = 10^{o}$
\cite{budagov96-72}
and obtained the values of $f_{\pi^{o}}$ are equal to $55 \%$ and $56 \%$
for Groom and Wigmans parameterisations of $f^{\prime}_{\pi^{o}}$,
respectively.
Thus, our measured value of $f_{\pi^o}$ agrees well with the experimental one 
\cite{acosta92} 
and Monte Carlo calculation 
\cite{gabrial94}
and with our calculations.

The determined contributions of parts for various depths give the
possibility to obtain the fractions of ``electromagnetic'' and 
``hadronic'' parts of hadronic showers in various stages 
of longitudinal development of showers.
On the Fig.\ \ref{fig:f15} (right)
shows the corresponding results.
As can be seen as shower developed the 
``electromagnetic'' fraction decreases.
This is natural since the shower energy exhausts and as a result 
$\pi^o$ production decreases.
The best fit to these data is $f_{em} = (76\pm2) - (8.1\pm0.4) \cdot x$.

\section{Conclusions}

We have investigated the
hadronic shower longitudinal and lateral profiles
on the basis of $100\ GeV$
pion beam data at incidence angle
$\Theta = 10^{o}$ at impact points $Z$ in the range from $- 36$ to $20$ cm.


\begin{itemize}
\item
Some useful formulae for investigating of lateral profiles 
have been derived:
        \begin{itemize}
        \item
        the integral expression (\ref{e5}) for radial density $\Phi (r)$
        as a function of marginal density $f  (z)$,
        \item
        the formula (\ref{e23}) for radial density $\Phi (r)$ 
        and the formula (16) and (\ref{e23-2}) for cumulative function $F (z)$
        in case of three exponential form of $f  (z)$ (\ref{e21}). 
        \end{itemize}
\item
We have obtained for four depths and for overall calorimeter:
        \begin{itemize}
        \item
        the energy depositions in towers, $E (z)$;
        \item
        the cumulative functions, $F (z)$;
        \item
        the marginal transversal densities, $f  (z)$;
        \item
        the underling radial energy densities, $\Phi (r)$;
        \item
        the containment of a shower as a function of radius, $I (r)$;
        \item
        the radii of cylinders for given shower containment;
        \item
        the fractions of ``electromagnetic'' part of a shower;
        \item
        the differential longitudinal energy deposition $\Delta E / \Delta x$; 
        \item
        the three-dimensional hadronic shower parametrisation.
        \end{itemize}
\end{itemize}

We have compared our data with relevant data for conventional 
iron-scintillator calorimeters, SPACAL lead-scintillating fiber calorimeter  
and Monte Carlo calculations.
Our longitudinal profile agree with the ones for a conventional 
iron-scintillator calorimeters and Monte Carlo prediction.
Our lateral profile is not agree with the Monte Carlo prediction.

The three-dimensional hadronic shower 
pa\-ra\-met\-ri\-za\-tion
for iron-scin\-til\-la\-tor calorimeter have been obtained.
This parametrisation is important in fast Monte-Carlo 
simulation for ATLAS calorimetry.

\section{Appendix 1.}{\large\bf Solution of Abelian equation.}
\bigskip

It is stated above that the marginal density distribution
$f (z)$ is connected with the radial energy density $\Phi(r)$ by
relation (\ref{e4}).
This integral equation can be reduced to the Abelian equation
\cite{whitteker27}.
Here we show how to solve the equation (\ref{e4}) and to obtain
the expression (\ref{e5}).
Let $\xi=r^{2}$ and $\eta=z^{2}$
so that the equation (\ref{e4}) becomes
\begin{equation}
f  (\sqrt{\eta}) = 
 \int_{ \eta }^{ \infty } 
\frac{ \Phi ( \sqrt{ \xi } ), d \xi }{ \sqrt( \xi - \eta ) }.
\label{e11}
\end{equation}

If we multiply (\ref{e11}) on $\sqrt{ \eta - \alpha }$
and obtained product integrate over $d\eta$ in limits $[ \alpha, \infty]$
then
\begin{equation}
I = 
\int_{\alpha}^{\infty} 
\frac{f  (\sqrt{\eta}) }{ \sqrt{ \eta - \alpha } }\, d \eta
\label{e12}
\end{equation}
Using the following relation
\begin{equation}
\int_{\alpha}^{\infty}d\eta\int_\eta^{\infty}d \xi \ldots =
\int_{\alpha}^{\infty}d\xi\int_\alpha^\xi d \eta \ldots
\label{e13}
\end{equation}
we get
\begin{equation}
I=\int_0^\infty\Phi(\sqrt{\xi}) d\xi\int_\alpha^\xi\frac{d\eta}
{\sqrt{(\xi-\eta)(\eta-\alpha)}}
\label{e14}
\end{equation}
We should be able to use the fact that
\begin{equation}
\int_\alpha^\xi \frac{d \eta}
{ \sqrt{ ( \xi - \eta )( \eta - \alpha ) }} = \pi
\label{e15}
\end{equation}
and write
\begin{equation}
I = \pi \int_{\alpha}^{\infty} \Phi ( \sqrt{ \xi } ) d \xi.
\label{e16}
\end{equation}
Let us return to former variables $z$,  $r$ and  $\alpha = r^2$.
Then (\ref{e12}) and (\ref{e16}) modified into
\begin{equation}
{\int}_{ r }^{\infty} 
\frac{z f(z)}{ \sqrt{z^2 - r^2} } dz =
\pi {\int}_{r}^{\infty} \Phi ( r ) r dr
\label{e17}
\end{equation}

Differentiating equation (\ref{e17}) over $r$ we get 
\begin{equation}
\Phi(r) = - \frac{1}{\pi} \frac{d}{dr^2}
\int_{r^2}^{\infty} \frac{f  (z) d z^2}{\sqrt(z^{2}-r^{2})}.
\label{e500}
\end{equation}

\section{Appendix 2.}

{\large\bf
 The effective nuclear interaction length,
the effective radiation length and the effective Moliere radius.}
\bigskip

In reality our calorimeter represents itself the complex
structure of various materials and it is necessary to know the effective
nuclear
interaction length ($\lambda^{eff}$), the effective radiation length
($X_{o}^{eff}$)
and the effective Moliere radius ($R_{M}^{eff}$).

We calculated these quantities for our calorimeter.
For calculating $\lambda^{eff}$ and $X_{o}^{eff}$ we used the
algorithm suggested in
\cite{lokajicek95-64} so,
\begin{equation}
\lambda^{eff} =
\frac{x_{Fe}+x_{Sc}+x_{W}+x_{A}}
{x_{Fe}/\lambda_{Fe}+x_{Sc}/\lambda_{Sc}+x_{W}/\lambda_{W}+x_{A}/\lambda_{A}},
\label{ap-21}
\end{equation}
where $x_{Fe}$, $x_{Sc}$, $x_{W}$, $x_{A}$ are the volume fractions of
the respective
materials
(Fe, scintillator, wrapping, air) in a period of 18 mm thick of the
calorimeter,
$\lambda_{i}$ are the corresponding interaction length.

The effective radiation length, $X_{o}^{eff}$, was also calculated by
the formula
(\ref{ap-21}) replacing the corresponding $\lambda_{i}$ values by
$X_{o,i}$ values.
The Moliere radii is equal \cite{review96}
\begin{equation}
R_M = X_o E_s / E_c
\label{ap-22}
\end{equation}
and for a mixture of compound
\begin{equation}
\frac{1}{R_M} = \frac{1}{E_s} \sum_{j}\frac{x_j E_{c}^{j}}{X_{o}^{j}},
\label{ap-23}
\end{equation}
where $x_j$ are the fraction by weight, $E_s = 21.2\ MeV$ is the scale energy,
$E_{c,j}$ are the critical energy.
Critical energy for the chemical elements with the atomic number
of $Z$ is equal
$E_c = 610\ MeV / (Z +1.24)$.
The values of the $x_j$, $\lambda_{j}$
and $X_{o}^{j}$ are given in Table \ref{tb-a21}

\begin{table}[tbph]
\caption{
\label{tb-a21}}
\begin{center}
\begin{tabular}{|l|c|c|c|}
\hline
Material  & $x_j$  & $\lambda_j$, mm     & $X_{o}^{j}$, mm\\
\hline
\hline
Fe        &  14/18 & 168             & 17.6         \\
\hline
Sc        &   3/18 & 795             & 424         \\
\hline
Wrapping  & 0.2/18 & $\lambda_{Sc}$ & $X_{o}^{Sc}$  \\
\hline
Air       & 0.8/18 & 747000            & 304200        \\
\hline
\end{tabular}
\end{center}
\end{table}

In the Table \ref{tb-a22} the results of the our calculations are given.
In the fourth column the corresponding values for SPACAL
\cite{acosta92}
are also shown for comparison.
The corresponding values for basic materials of
these calorimeters (Fe, Pb) from
\cite{review96} are also given.

\begin{table}[tbph]
\caption{
\label{tb-a22}}
\begin{center}
\begin{tabular}{|l|c|c|c|c|}
\hline
 & \multicolumn{2}{|c|}{TILECAL} & \multicolumn{2}{|c|}{SPACAL}  \\
\cline{2-5}
 & eff, mm & Fe, mm  & eff, mm  & Pb, mm  \\
\hline
\hline
$X_{o}$         & 22.4 & 17.6 & 7.2  & 5.6  \\
\hline
$R_{M}$         & 20.5 & 16.6 & 20.  & 16.3 \\
\hline
$\lambda_p$     & 206. & 168. & 210. & 171. \\
\hline
$\lambda_\pi$   & 251. & 207. & 244. & 198. \\
\hline
\end{tabular}
\end{center}
\end{table}

It turns out that the values $\lambda^{eff}$ and $R_{M}^{eff}$ for these two
calorimeters are approximately equal.
The effective nuclear interaction length for pion was also calculated
by using relation  $\lambda_{\pi} / \lambda_{p}$ (Appendix 3).

\section{Appendix 3.}

{\large\bf
The nuclear interaction length for pion.}
\bigskip

Usually in hadronic 
calorimetry the nuclear interaction length (${\lambda}_{I}$)
given in Review of Particle Physics \cite{review96} is used.
It is the mean free path for protons between inelastic interactions,
calculated using the expression
\begin{equation}
{\lambda}_{I} = A / (N_{A} {\sigma}_{I} \rho) ,
\label{ap-1}
\end{equation}
where $A$ is atomic weight,
$N_{A}$ is Avogadro number,
${\sigma}_{I}$ is the nuclear inelastic cross section,
$\rho$ is a density.
For $Fe$ (${\sigma}_{I} = 703\ mb$) it amount to 168 mm.
But sometimes the pion interaction length is needed as in our case.
We calculated the pion and proton interaction lengths and compared with
\cite{review96} and with the some experimental data
\cite{holder78}, \cite{huges90}.
For this we used the absorption cross section of pion and proton on nuclei in
the range $60 \div 280$ $GeV/c$ measured by \cite{carroll79}.
Unfortunately in this  work absorption cross section of $\pi$ and  $p$
on $Fe$ nuclei are not measured.
Therefore we used the nearest cross section for $Cu$ (see Table~\ref{at-1}).
Transformation from $Cu$ to $Fe$ was performed using the expression
\begin{equation}
{\sigma}_{Fe} = {\sigma}_{Cu} {(A_{Fe} / A_{Cu})}^{\alpha}.
\label{ap-2}
\end{equation}

\begin{table}[tbph]
\caption{
Measured absorption cross sections and parameter $\alpha$ for $Cu$ target.
\label{at-1}}
\begin{center}
\begin{tabular}{|c|c|c|c|c|}
\hline
E & \multicolumn{2}{|c|}{60 GeV} & \multicolumn{2}{|c|}{200 GeV}  \\
\hline
\hline
  & $\sigma$, mb & $\alpha$ &  $\sigma$, mb & $\alpha$  \\
\hline
$\pi^+$ & 627 $\pm$ 19 & 0.764 $\pm$ 0.01 & 629 $\pm$ 19 & 0.762 $\pm$ 0.01 \\
\hline
  $p^+$ & 764 $\pm$ 23 & 0.719 $\pm$ 0.01 & 774 $\pm$ 23 & 0.719 $\pm$ 0.01 \\
\hline
$\lambda_{\pi} / \lambda_{p}$&
\multicolumn{2}{|c|}{$1.22 \pm 0.05$} &
\multicolumn{2}{|c|}{$1.23 \pm 0.05$}  \\
\hline
$\alpha_{p} - \alpha_{\pi}$&
\multicolumn{2}{|c|}{$- 0.045 \pm 0.014$} &
\multicolumn{2}{|c|}{$- 0.043 \pm 0.014$} \\
\hline
\end{tabular}
\end{center}
\end{table}

In Table~\ref{at-1} the measured cross sections for $\pi^+ Cu$
interactions at 60 and 200 $GeV$ together with parameter $\alpha$
are given.
Using expressions (\ref{ap-1}) and (\ref{ap-2}) cross sections
and nuclei interaction lengths are presented in Table~\ref{at-2}.

As can be seen from Table~\ref{at-2} the $Fe$ cross sections and nuclear
interaction lengths do not depend from energy within errors at
$60 \div 200$.
The mean over energy range values are
${\lambda}_{\pi} = 207 \pm 7$ mm, ${\lambda}_{p} = 168 \pm 5$ mm,
${\lambda}_{\pi} / {\lambda}_{p} = 1.22 \pm 0.05$.
The value of ${\lambda}_{p}$ coincides with that given in \cite{review96}.
The value of ${\lambda}_{\pi}$ may be compared with the measured in
\cite{huges90} value ${\lambda}_{\pi} = 200 \pm 10$ mm
and in \cite{holder78} value ${\lambda}_{\pi} = 190 $ mm.

\begin{table}[tbph]
\caption{
Calculated cross sections and nuclei interaction lengths for $Fe$ target.
\label{at-2}}
\begin{center}
\begin{tabular}{|c|c|c|}
\hline
E       & 60  GeV &  200 GeV   \\
\hline
\hline
${\sigma}_{\pi}$, mb & 568 $\pm$ 18 & 570 $\pm$ 18 \\
\hline
${\sigma}_{p}$, mb   & 696 $\pm$ 21 & 705 $\pm$ 23 \\
\hline
${\lambda}_{\pi}$, mm& 207 $\pm$ 7 & 207 $\pm$ 7 \\
\hline
${\lambda}_{p}$, mm  & 169 $\pm$ 5 & 167 $\pm$ 5 \\
\hline
${\lambda}_{\pi} / {\lambda}_{p}$
                     & 1.22 $\pm$ 0.05 & 1.24 $\pm$ 0.05 \\
\hline
\end{tabular}
\end{center}
\end{table}

The value measured in \cite{huges90}  presumably have be corrected
on iron equivalent length (including scintillators) and then amounts to
$210 \pm 10$ mm and better agrees with ours.

Using the expressions (\ref{ap-1}) and (\ref{ap-2})
 we obtained the $A$-dependence
of the ratio
\begin{equation}
{\lambda}_{\pi}  / {\lambda}_{p} (A) =
{\lambda}_{\pi}  / {\lambda}_{p} (Cu)
{(A / A_{Cu})}^{{\alpha}_{p} - {\alpha}_{\pi}} ,
\label{ap-3}
\end{equation}
where ${\alpha}_{p} - {\alpha}_{\pi} = 0.045 \pm 0.014$
(Table~\ref{at-1}).
For $Pb$ nucleus it corresponds to
${\lambda}_{\pi}  / {\lambda}_{p} (Pb) = 1.16 \pm 0.05$.
Therefore, it is mistaken to use the value of the ratio
${\lambda}_{\pi}  / {\lambda}_{p} (Pb) = 1.5$ as have been made in
\cite{acosta91} when analysing of data from lead scintillating
fiber spaghetti calorimeter.

\section{Acknowledgements}

This work is the result of the efforts of many people from ATLAS
Collaboration.
The authors are greatly indebted to all Collaboration
for their test beam setup and data taking.

Authors are grateful Peter Jenni for his attention
and support of this work.
We are indebted to 
M.\ Cavalli-Sforza, 
M.\ Bosman, 
I.\ Efthymiopoulos, 
A.\ Henriques, 
B.\ Stanek 
and
I.\ Vichou 
for
the valuable discussions.
We are thankful to S.\ Hellman  for the
careful reading and constructive advices on the improvement
of paper context.






}
\end{document}